\input harvmac
\input epsf

\overfullrule=0pt
\abovedisplayskip=12pt plus 3pt minus 3pt
\belowdisplayskip=12pt plus 3pt minus 3pt
%
%
\font\zfont = cmss10 
\font\litfont = cmr6

%

\def\tilde{\widetilde}
\def\bar{\overline}
\def\to{\rightarrow}

\def\cN{{\cal N}}

\def\bigone{\hbox{1\kern -.23em {\rm l}}}
\def\ZZ{\hbox{\zfont Z\kern-.4emZ}}
\def\half{{\litfont {1 \over 2}}}

\font\litfont=cmr6

\def\Dbar{${\bar {\rm D}}$}

\font\zfont = cmss10 
\font\litfont = cmr6

\def\bigone{\hbox{1\kern -.23em {\rm l}}}
\def\ZZ{\hbox{\zfont Z\kern-.4emZ}}
\def\half{{\litfont {1 \over 2}}}




\def\frac#1#2{{#1 \over #2}}

\def\cite#1{\#1}

\def\section#1{\newsec {#1}}
\def\subsection#1{\subsec {#1}}
\def\subsubsection#1{\bigskip\noindent{\it #1}}


\def\myI#1 {\int \! #1 \,}

\def\eqn#1#2{\xdef #1{(\secsym\the\meqno)}\writedef{#1\leftbracket#1}%
    $$#2\eqno#1\eqlabeL#1$$%
    \xdef #1{(eq.~\secsym\the\meqno) }\global\advance\meqno by1}
%



\def\paren#1{\left( #1 \right)}


\def\comm#1#2{\left[ #1, #2 \right]}

\def\pa{\partial}

\def\cA{{\cal A}}
\def\cF{{\cal F}}




%
\let\useblackboard=\iftrue
%
%
\newfam\black
\def\Title#1#2{\rightline{#1}
\ifx\answ\bigans\nopagenumbers\pageno0\vskip1in%
\baselineskip 15pt plus 1pt minus 1pt
\else
\def\listrefs{\footatend\vskip 1in\immediate\closeout\rfile\writestoppt
\baselineskip=14pt\centerline{{\bf References}}\bigskip{\frenchspacing%
\parindent=20pt\escapechar=` \input
refs.tmp\vfill\eject}\nonfrenchspacing}
\pageno1\vskip.8in\fi \centerline{\titlefont #2}\vskip .5in}

\ifx\answ\bigans\def\tcbreak#1{}\else\def\tcbreak#1{\cr&{#1}}\fi
\useblackboard
\message{If you do not have msbm (blackboard bold) fonts,}
\message{change the option at the top of the tex file.}
\font\blackboard=msbm10 scaled \magstep1
\font\blackboards=msbm7
\font\blackboardss=msbm5
\textfont\black=\blackboard
\scriptfont\black=\blackboards
\scriptscriptfont\black=\blackboardss

\else

\fi



%
\message{S-Tables Macro v1.0, ACS, TAMU (RANHELP@VENUS.TAMU.EDU)}
%
%
\newhelp\stablestylehelp{You must choose a style between 0 and 3.}%
\newhelp\stablelinehelp{You should not use special hrules when stretching
a table.}%
\newhelp\stablesmultiplehelp{You have tried to place an S-Table
inside another S-Table.  I would recommend not going on.}%
%
%
\newdimen\stablesthinline
\stablesthinline=0.4pt
\newdimen\stablesthickline
\stablesthickline=1pt
%
%
\newif\ifstablesborderthin
\stablesborderthinfalse
\newif\ifstablesinternalthin
\stablesinternalthintrue
\newif\ifstablesomit
\newif\ifstablemode
\newif\ifstablesright
\stablesrightfalse
%
%
\newdimen\stablesbaselineskip
\newdimen\stableslineskip
\newdimen\stableslineskiplimit
%
%
\newcount\stablesmode
\newcount\stableslines
\newcount\stablestemp
\stablestemp=3
\newcount\stablescount
\stablescount=0
\newcount\stableslinet
\stableslinet=0
%
%
%
\newcount\stablestyle
\stablestyle=0
%
%
\def\stablesleft{\quad\hfil}%
\def\stablesright{\hfil\quad}%
%
%
\catcode`\|=\active%
%
%
\newcount\stablestrutsize
\newbox\stablestrutbox
\setbox\stablestrutbox=\hbox{\vrule height10pt depth5pt width0pt}
\def\stablestrut{\relax\ifmmode%
                         \copy\stablestrutbox%
                       \else%
                         \unhcopy\stablestrutbox%
                       \fi}%
%
%
\newdimen\stablesborderwidth
\newdimen\stablesinternalwidth
\newdimen\stablesdummy
\newcount\stablesdummyc
\newif\ifstablesin
\stablesinfalse
%
%
\def\begintable{\stablestart%
  \stablemodetrue%
  \stablesadj%
  \halign%
  \stablesdef}%
\def\stablesadj{%
  \ifcase\stablestyle%
    \hbox to \hsize\bgroup\hss\vbox\bgroup%
  \or%
    \hbox to \hsize\bgroup\vbox\bgroup%
  \or%
    \hbox to \hsize\bgroup\hss\vbox\bgroup%
  \or%
    \hbox\bgroup\vbox\bgroup%
  \else%
    \errhelp=\stablestylehelp%
    \errmessage{Invalid style selected, using default}%
    \hbox to \hsize\bgroup\hss\vbox\bgroup%
  \fi}%
\def\stablesend{\egroup%
  \ifcase\stablestyle%
    \hss\egroup%
  \or%
    \hss\egroup%
  \or%
    \egroup%
  \or%
    \egroup%
  \else%
    \hss\egroup%
  \fi}%
\def\stablestart{%
  \ifstablesin%
    \errhelp=\stablesmultiplehelp%
    \errmessage{An S-Table cannot be placed within an S-Table!}%
  \fi
  \global\stablesintrue%
  \global\advance\stablescount by 1%
  \message{<S-Tables Generating Table \number\stablescount}%
  \begingroup%
  \stablestrutsize=\ht\stablestrutbox%
  \advance\stablestrutsize by \dp\stablestrutbox%
  \ifstablesborderthin%
    \stablesborderwidth=\stablesthinline%
  \else%
    \stablesborderwidth=\stablesthickline%
  \fi%
  \ifstablesinternalthin%
    \stablesinternalwidth=\stablesthinline%
  \else%
    \stablesinternalwidth=\stablesthickline%
  \fi%
  \tabskip=0pt%
  \stablesbaselineskip=\baselineskip%
  \stableslineskip=\lineskip%
  \stableslineskiplimit=\lineskiplimit%
  \offinterlineskip%
  \def\borderrule{\vrule width \stablesborderwidth}%
  \def\internalrule{\vrule width \stablesinternalwidth}%
  \def\thinline{\noalign{\hrule height \stablesthinline}}%
  \def\thickline{\noalign{\hrule height \stablesthickline}}%
  \def\trule{\omit\leaders\hrule height \stablesthinline\hfill}%
  \def\ttrule{\omit\leaders\hrule height \stablesthickline\hfill}%
  \def\tttrule##1{\omit\leaders\hrule height ##1\hfill}%
  \def\stablesel{&\omit\global\stablesmode=0%
    \global\advance\stableslines by 1\borderrule\hfil\cr}%
  \def\el{\stablesel&}%
  \def\elt{\stablesel\thinline&}%
  \def\eltt{\stablesel\thickline&}%
  \def\elttt##1{\stablesel\noalign{\hrule height ##1}&}%
  \def\elspec{&\omit\hfil\borderrule\cr\omit\borderrule&%
              \ifstablemode%
              \else%
                \errhelp=\stablelinehelp%
                \errmessage{Special ruling will not display properly}%
              \fi}%
  \def\stmultispan##1{\mscount=##1 \loop\ifnum\mscount>3 \stspan\repeat}%
  \def\stspan{\span\omit \advance\mscount by -1}%
  \def\multicolumn##1{\omit\multiply\stablestemp by ##1%
     \stmultispan{\stablestemp}%
     \advance\stablesmode by ##1%
     \advance\stablesmode by -1%
     \stablestemp=3}%
  \def\multirow##1{\stablesdummyc=##1\parindent=0pt\setbox0\hbox\bgroup%
    \aftergroup\emultirow\let\temp=}
  \def\emultirow{\setbox1\vbox to\stablesdummyc\stablestrutsize%
    {\hsize\wd0\vfil\box0\vfil}%
    \ht1=\ht\stablestrutbox%
    \dp1=\dp\stablestrutbox%
    \box1}%
  \def\stpar##1{\vtop\bgroup\hsize ##1%
     \baselineskip=\stablesbaselineskip%
     \lineskip=\stableslineskip%
   \lineskiplimit=\stableslineskiplimit\bgroup\aftergroup\estpar\let\temp=}%
  \def\estpar{\vskip 6pt\egroup}%
  \def\stparrow##1##2{\stablesdummy=##2%
     \setbox0=\vtop to ##1\stablestrutsize\bgroup%
     \hsize\stablesdummy%
     \baselineskip=\stablesbaselineskip%
     \lineskip=\stableslineskip%
     \lineskiplimit=\stableslineskiplimit%
     \bgroup\vfil\aftergroup\estparrow%
     \let\temp=}%
  \def\estparrow{\vfil\egroup%
     \ht0=\ht\stablestrutbox%
     \dp0=\dp\stablestrutbox%
     \wd0=\stablesdummy%
     \box0}%
  \def|{\global\advance\stablesmode by 1&&&}%
  \def\|{\global\advance\stablesmode by 1&\omit\vrule width 0pt%
         \hfil&&}%
\def\vt{\global\advance\stablesmode
by 1&\omit\vrule width \stablesthinline%
          \hfil&&}%
  \def\vtt{\global\advance\stablesmode by 1&\omit\vrule width
\stablesthickline%
          \hfil&&}%
  \def\vttt##1{\global\advance\stablesmode by 1&\omit\vrule width ##1%
          \hfil&&}%
  \def\vtr{\global\advance\stablesmode by 1&\omit\hfil\vrule width%
           \stablesthinline&&}%
  \def\vttr{\global\advance\stablesmode by 1&\omit\hfil\vrule width%
            \stablesthickline&&}%
\def\vtttr##1{\global\advance\stablesmode
 by 1&\omit\hfil\vrule width ##1&&}%
  \stableslines=0%
  \stablesomitfalse}
\def\stablesdef{\bgroup\stablestrut\borderrule##\tabskip=0pt plus 1fil%
  &\stablesleft##\stablesright%
  &##\ifstablesright\hfill\fi\internalrule\ifstablesright\else\hfill\fi%
  \tabskip 0pt&&##\hfil\tabskip=0pt plus 1fil%
  &\stablesleft##\stablesright%
  &##\ifstablesright\hfill\fi\internalrule\ifstablesright\else\hfill\fi%
  \tabskip=0pt\cr%
  \ifstablesborderthin%
    \thinline%
  \else%
    \thickline%
  \fi&%
}%
\def\endtable{\advance\stableslines by 1\advance\stablesmode by 1%
   \message{- Rows: \number\stableslines, Columns:  \number\stablesmode>}%
   \stablesel%
   \ifstablesborderthin%
     \thinline%
   \else%
     \thickline%
   \fi%
   \egroup\stablesend%
\endgroup%
\global\stablesinfalse}
%



\Title{\vtop{\hbox{hep-th/0306030}
\hbox{SU-ITP-03/11, SLAC-PUB-9907}}}
{\vbox{\centerline{On Branes and Oriented B-fields}}}
\centerline{{\bf Keshav Dasgupta}\foot{\tt keshav@itp.stanford.edu}~~ and ~~
{\bf Marina Shmakova}\foot{\tt shmakova@slac.stanford.edu}}
\vskip 15pt
\centerline{\it ${}^1$Varian Laboratory of Physics, Stanford University,
Stanford CA 94305-4060, USA}
\vskip 5pt
\centerline{\it ${}^2$Stanford Linear Accelerator Center, Stanford University,
Stanford CA 94309, USA}

\ \smallskip
\centerline{\bf Abstract}

\noindent Novel theories appear on the world-volume of branes
by orienting $B$ fields along various directions of the branes. We review
some of the earlier developments and explore many
new examples of these theories. In
particular, among other things,
we study the pinning effect of branes near conifold like
singularities and brane-antibrane theories with different fluxes on their
world-volumes. We show that all these
 theories arise from different limits of an M-theory configuration
with appropriately chosen G-fluxes. This gives us a way to study them
from a unified framework in M-theory.

\Date{May 2003}


{\vfill\eject}
\ftno=0

\lref\CDS{A.~Connes, M.R.~Douglas and A.~Schwarz,
  {\it ``Noncommutative Geometry and Matrix Theory:
  Compactification on Tori,''} JHEP {\bf 9802} (1998) 003, hep-th/9711162.}

\lref\DH{M.R.~Douglas and C.~Hull,
  {\it ``D-branes and the Noncommutative Torus,''}
  JHEP {\bf 9802} (1998) 008, hep-th/9711165.}

\lref\seiwit{N. Seiberg and E. Witten,
  {\it ``String Theory and Noncommutative Geometry,''}
  JHEP {\bf 9909}, 032 (1999, hep-th/9908142.}

\lref\CDGR{S. Chakravarty, K. Dasgupta,
   O.J. Ganor and G. Rajesh,
    {\it ``Pinned Branes and New Non-Lorentz Invariant Theories,''}
   Nucl.Phys. {\bf B587} (2000) 228, hep-th/ 0002175.}

\lref\ZYIN{Z. Yin, {\it ``A Note on Space Noncommutativity,''}
   Phys. Lett. {\bf B466} (1999) 234, hep-th/9908152.}

\lref\berraj{K. Dasgupta, O. J. Ganor and G. Rajesh,
  {\it ``Vector Deformations of $\cN = 4$ Yang-Mills Theory, Pinned Branes
 and Arched Strings,''} JHEP {\bf 0104}, 034 (2001), hep-th/0010072.}

\lref\DolNap{L.~Dolan and C.~R.~Nappi,
{\it ``A scaling limit with many noncommutativity parameters,''}
Phys.\ Lett.\ B {\bf 504}, 329 (2001), hep-th/0009225.}

\lref\tataroh{R.~de Mello Koch, K.~Oh and R.~Tatar,
{\it ``Moduli space for conifolds as intersection of orthogonal D6 branes,''}
Nucl.\ Phys.\ B {\bf 555}, 457 (1999), hep-th/9812097.}

\lref\alisha{M.~Alishahiha, Y.~Oz and M.~M.~Sheikh-Jabbari,
{\it ``Supergravity and large N noncommutative field theories,''}
JHEP {\bf 9911}, 007 (1999), hep-th/9909215.}

\lref\geometric{J.~D.~Edelstein, K.~Oh and R.~Tatar,
{\it ``Orientifold, geometric transition and large N 
duality for SO/Sp gauge  theories,''}
JHEP {\bf 0105}, 009 (2001), hep-th/0104037;
K.~Dasgupta, K.~Oh and R.~Tatar,
{\it ``Geometric transition, large N dualities and MQCD dynamics,''}
Nucl.\ Phys.\ B {\bf 610}, 331 (2001), hep-th/0105066;
{\it ``Open/closed string dualities and Seiberg
duality from geometric  transitions in M-theory,''}
JHEP {\bf 0208}, 026 (2002), hep-th/0106040;
K.~Dasgupta, K.~h.~Oh, J.~Park and R.~Tatar,
{\it ``Geometric transition versus cascading solution,''}
JHEP {\bf 0201}, 031 (2002), hep-th/0110050;
A.~Giveon, A.~Kehagias and H.~Partouche,
{\it ``Geometric transitions, brane dynamics and gauge theories,''}
JHEP {\bf 0112}, 021 (2001), hep-th/0110115;
K.~h.~Oh and R.~Tatar,
{\it ``Duality and confinement in N = 1 supersymmetric 
theories from geometric  transitions,''}
Adv.\ Theor.\ Math.\ Phys.\  {\bf 6}, 141 (2003), hep-th/0112040.}

\lref\bbdg{K.~Becker, M.~Becker, K.~Dasgupta and P.~S.~Green,
{\it ``Compactifications of heterotic theory on
non-Kaehler complex manifolds.  I,''}
JHEP {\bf 0304}, 007 (2003), hep-th/0301161.}

\lref\ceqlone{E.~Witten,
{\it ``Ground ring of two-dimensional string theory,''}
Nucl.\ Phys.\ B {\bf 373}, 187 (1992), hep-th/9108004;
E.~Witten and B.~Zwiebach,
{\it ``Algebraic structures and differential geometry in $2-D$
string theory,''}
Nucl.\ Phys.\ B {\bf 377}, 55 (1992), hep-th/9201056.}

\lref\ghomuk{D.~Ghoshal, D.~P.~Jatkar and S.~Mukhi,
{\it ``Kleinian singularities and the ground ring of C=1 string theory,''}
Nucl.\ Phys.\ B {\bf 395}, 144 (1993), hep-th/9206080.}

\lref\shahin{
M.~M.~Sheikh-Jabbari,
{\it ``Noncommutative super Yang-Mills theories
with 8 supercharges and brane  configurations,''}
Nucl.\ Phys.\ B {\bf 587}, 195 (2000), hep-th/0001089.}

\lref\RT {R. Tatar, {\it ``A Note on Noncommutative Field Theory and
  Stability of Brane- Antibrane Systems,''}
 hep-th/0009213.}

\lref\senm{A.~Sen,
{\it ``Dynamics of multiple Kaluza-Klein monopoles in M and string theory,''}
Adv.\ Theor.\ Math.\ Phys.\  {\bf 1}, 115 (1998), hep-th/9707042;
{\it ``A note on enhanced gauge symmetries in M- and string theory,''}
JHEP {\bf 9709}, 001 (1997), hep-th/9707123;
{\it ``Strong coupling dynamics of branes from M-theory,''}
JHEP {\bf 9710}, 002 (1997), hep-th/9708002.}

\lref\dasmuk{S.~Mukhi, {\it Private Communications};
K.~Dasgupta and S.~Mukhi, {\it Unpublished Notes} (1999).}

\lref\renata{K.~Dasgupta, C.~Herdeiro, S.~Hirano and R.~Kallosh,
{\it ``D3/D7 inflationary model and M-theory,''}
Phys.\ Rev.\ D {\bf 65}, 126002 (2002), hep-th/0203019.}

\lref\DM{K.~Dasgupta and S.~Mukhi,
{\it ``Brane constructions, conifolds and M-theory,''}
Nucl.\ Phys.\ B {\bf 551}, 204 (1999), hep-th/9811139.}

\lref\karch{A.~Hanany and A.~M.~Uranga,
{\it ``Brane boxes and branes on singularities,''}
JHEP {\bf 9805}, 013 (1998), hep-th/9805139;
M.~Aganagic, A.~Karch, D.~Lust and A.~Miemiec,
{\it ``Mirror symmetries for brane configurations
and branes at singularities,''}
Nucl.\ Phys.\ B {\bf 569}, 277 (2000), hep-th/9903093.}

\lref\corshi{L.~Cornalba and R.~Schiappa,
{\it ``Nonassociative star product deformations for D-brane
worldvolumes in  curved backgrounds,''}
Commun.\ Math.\ Phys.\  {\bf 225}, 33 (2002), hep-th/0101219.}

\lref\angle{
A.~M.~Uranga,
{\it ``Brane configurations for branes at conifolds,''}
JHEP {\bf 9901}, 022 (1999), hep-th/9811004.}

\lref\difuli{M.~Alishahiha and H.~Yavartanoo,
{\it ``Supergravity description of the
large N noncommutative dipole field  theories,''}
JHEP {\bf 0204}, 031 (2002), hep-th/0202131;
N.~Sadooghi and M.~Soroush,
{\it ``Noncommutative dipole QED,''}
Int.\ J.\ Mod.\ Phys.\ A {\bf 18}, 97 (2003), hep-th/0206009.}

\lref\difnew{O.~J.~Ganor and U.~Varadarajan,
{\it ``Nonlocal effects on D-branes in plane-wave backgrounds,''}
JHEP {\bf 0211}, 051 (2002), hep-th/0210035;
M.~Alishahiha and O.~J.~Ganor,
{\it ``Twisted backgrounds, pp-waves and nonlocal field theories,''}
JHEP {\bf 0303}, 006 (2003), hep-th/0301080;
M.~Alishahiha and A.~E.~Mosaffa,
{\it ``Semiclassical string solutions on deformed NS5-brane
backgrounds and  new pp-wave,''} hep-th/0302005.}

\lref\sadovafa{M.~Bershadsky, C.~Vafa and V.~Sadov,
{\it ``D-Strings on D-Manifolds,''}
Nucl.\ Phys.\ B {\bf 463}, 398 (1996), hep-th/9510225;
H.~Ooguri and C.~Vafa,
{\it ``Two-Dimensional Black Hole and Singularities of CY Manifolds,''}
Nucl.\ Phys.\ B {\bf 463}, 55 (1996), hep-th/9511164.}

\lref\bak{D.~s.~Bak and N.~Ohta,
{\it ``Supersymmetric D2 anti-D2 strings,''}
Phys.\ Lett.\ B {\bf 527}, 131 (2002), hep-th/0112034.}

\lref\WITSFT {E. Witten, {\it ``Noncommutative Geometry and String Field
  Theory,''} Nucl. Phys. {\bf B268} (1986) 253.}

\lref\jabdip{M.~M.~Sheikh-Jabbari,
{\it ``More on mixed boundary conditions and D-branes bound states,''}
Phys.\ Lett.\ B {\bf 425}, 48 (1998), hep-th/9712199;
{\it ``Open strings in a B-field background as electric dipoles,''}
Phys.\ Lett.\ B {\bf 455}, 129 (1999), hep-th/9901080;
D.~Bigatti and L.~Susskind,
{\it ``Magnetic fields, branes and noncommutative geometry,''}
Phys.\ Rev.\ D {\bf 62}, 066004 (2000), hep-th/9908056.}

\lref\CHU{C.-S. Chu and P.-M. Ho, {\it ``Constrained Quantization of Open
 String in Background $B$-Field and Noncommutative D-Brane,''}
Nucl. Phys. {\bf B568} (2000) 447, hep-th/9906192.}

\lref\KONTS {M. Kontsevich, {\it ``Deformation Quantization of Poisson
 Manifolds,''} q-alg/9709040.}

\lref\CandF {A. S. Cattaneo and  G. Felder, {\it ``A Path Integral Approach
 to Kontsevich Quantization Formula,''}
  Commun. Math. Phys. {\bf 212} (2000) 591, math.qa/9902090.}

\lref\DGS{ K. Dasgupta, G. Rajesh and S. Sethi, {\it ``M-Theory,
 Orientifolds and G-Flux,''} JHEP {\bf 9908} (1999) 023, hep-th/9908088.}

\lref\SJ{F. Arfaei and  M.M. Sheikh-Jabbari, {\it ``Dirac Quantization of Open
 Strings and Noncommutativity in Branes,''}
 Nucl. Phys. {\bf B576} (2000) 578, hep-th/9906161;
 M.M. Sheikh-Jabbari and A. Shirzad, {\it ``Boundary Conditions as Dirac
 Constraints,''} Eur.\ Phys.\ J.\ C {\bf 19}, 383 (2001), hep-th/9907055.}

\lref\KesYin{K. Dasgupta and Z. Yin, {\it ``Non-{\cA}belian Geometry''},
Commun.\ Math.\ Phys.\  {\bf 235}, 313 (2003),
hep-th/0011034.}

\lref\DMUK{K. Dasgupta and S. Mukhi, {\it ``Brane Constructions, Fractional
Branes and Anti- deSitter Domain Walls''}, JHEP {\bf 9907} (1999) 008,
hep-th/9904131.}

\lref\bergan{A.~Bergman and O.~J.~Ganor,
{\it ``Dipoles, twists and noncommutative gauge theory,''}
JHEP {\bf 0010}, 018 (2000), hep-th/0008030.}

\lref\savhashi{A.~Hashimoto and S.~Sethi,
{\it ``Holography and string dynamics in time-dependent backgrounds,''}
Phys.\ Rev.\ Lett.\  {\bf 89}, 261601 (2002), hep-th/0208126.}

\lref\hashi{A.~Hashimoto and N.~Itzhaki,
{\it ``Non-commutative Yang-Mills and the AdS/CFT correspondence,''}
Phys.\ Lett.\ B {\bf 465}, 142 (1999), hep-th/9907166;
J.~M.~Maldacena and J.~G.~Russo,
{\it ``Large N limit of non-commutative gauge theories,''}
JHEP {\bf 9909}, 025 (1999), hep-th/9908134.}

\lref\berdas{A.~Bergman, K.~Dasgupta, O.~J.~Ganor,
J.~L.~Karczmarek and G.~Rajesh,
{\it ``Nonlocal field theories and their gravity duals,''}
Phys.\ Rev.\ D {\bf 65}, 066005 (2002), hep-th/0103090.}

\lref\jabsad{D.~Sadri and M.~M.~Sheikh-Jabbari,
{\it ``String theory on parallelizable pp-waves,''} JHEP {\bf 0306}, 005 (2003),
hep-th/0304169.}

\lref\dasjab{K.~Dasgupta and M.~M.~Sheikh-Jabbari,
{\it ``Noncommutative dipole field theories,''}
JHEP {\bf 0202}, 002 (2002), hep-th/0112064.}

\lref\cita{K.~Dasgupta, G.~Rajesh, D.~Robbins and S.~Sethi,
{\it ``Time-dependent warping, fluxes, and NCYM,''}
JHEP {\bf 0303}, 041 (2003), hep-th/0302049.}

\lref\nekra{N.~Nekrasov and A.~Schwarz,
{\it ``Instantons on noncommutative R**4
and (2,0) superconformal six  dimensional theory,''}
Commun.\ Math.\ Phys.\  {\bf 198}, 689 (1998), hep-th/9802068.}

\listtoc
\writetoc

\section{Introduction}

Some time back theories on a M5 brane with lower
dimensional poincare invariance were considered. These theories suggested
a way to localise branes in ambient space. To generalise this further
we consider two cases. In one case, which
was studied earlier in some detail \refs\CDGR,
we keep a M5 brane near a single
centred Taub-NUT space. When a Lorentz noninvariant tensor background
is switched on we get a $(1,0)$ theory with a {\it massive} hypermultiplet
in six dimensions. This theory actually has a $4+1$ dimensional Poincare
symmetry. In the other case we keep the M5 brane near a conifold like
singularity. When a tensor background is switched on, we get a theory
with only $3+1$ dimensional poincare invariance. However, as we shall see,
this lowered poincare invariance is a consequence of the geometry of the
problem and not of the background tensor field.

The above discussion is only a small part of a more unified story that we want
to present here. It was studied earlier that switching on anti-symmetric
two form fluxes $-$ the usual $B_{NS}$ fluxes of string theory $-$ on the
brane results in making the coordinates non-commutative. This gave rise to the
non-commutative geometry as being an embedding of string theory. The
noncommutative geometry was elaborated in subsequent papers
and many interesting details were shown to occur. It was shown that the
dynamics are governed by a star product. In terms of the star product, both
the fields {\it and} the coordinates are noncommutative.

The noncommutative theory is also nonlocal in nature. However this is more
involved. There do exist a slightly simpler setup where the noncommutativity
only extends to the fields and not to the coordinates. The coordinates remain
commutative. This theory is also nonlocal by construction and is called the
{\it Dipole} theory, first constructed by Bergman and Ganor \refs\bergan.
Whereas the
noncommutative theory can have a maximal supersymmetry of ${\cal N} = 4$, the
dipole theory on the other hand can have a maximal susy of ${\cal N} = 2$.
Both the dipole and the noncommutative theory have supergravity duals whose
boundary metric degenerates along some of the directions. These details
have been explicitly shown in \refs\hashi\
 for the noncommutative geometry and in
\refs\berdas\ for the dipole theory. Since the coordinates
commute, the star product for the dipole theory is given by
\eqn\dipstar{ \Phi(x) \star \Psi(x) = e^{{1\over 2}(L_1 \cdot
{\del\over \del y} -L_2 \cdot {\del\over \del
w})}~\Phi(w)~\Psi(y)\vert^{y= w = x}} where $L_{1,2}$ are the
dipole lengths of the fields $\Phi$ and $\Psi$ respectively. This
star product is associative. As been discussed in \refs\berraj\
the dipole theory is derived from ${\cal N} = 4$ SYM theory by the
inclusion of a dimension five operator. This is again simpler than
noncommutative theory, where we need a dimension six operator.

An immediate extension of the above scenario is to consider the
non-abelian case by putting many branes parallel to each other.
There are two different scenarios now: (a) We have same $B$ fields
on all the parallel branes, and (b) Different branes have
different, but constant, $B$ fields. The first case is just a
simple extension of the Moyal star product in which the
non-abelian nature do not mix in any way with the
noncommutativity. The second case is more interesting as been
shown in \refs{\KesYin}. One could in fact evaluate the star
product which {\it mixes} both the non-abelian nature and the
noncommutativity of the branes in a unified way to give the
following nonabelian product: \eqn\nonabelly{ (\Phi\times
\Psi)^i_k (X)\equiv \sum_j~ \left( \Phi^i_j \star_{ijk} \Psi^j_k
\right) (X),} where $i,j,k$ label the branes and $\star_{ijk}$
depends on all three indices. This star product is associative and
the derivation of this has been given in details in
\refs{\KesYin}. In this paper we will elaborate some more
properties of this star product and discuss a situation where
there is a stable configuration.

All these configurations discussed above and in earlier papers
give us a setup which we call branes with {\it oriented} $B$
fields. Imagine a brane located at the degenerating point of a
Taub-NUT space with a $B$ flux switched on. The Taub-NUT is
required to have non-trivial configuration of $B$ fields so that
we cannot gauge it away. Following four interesting cases arise
now:

\item{1.} $B$ field is completely orthogonal to the brane. This is the
{\bf pinned brane} scenario where the brane is pinned at the origin of the
Taub-NUT. The hypermultiplets in this theory become massive and therefore the
susy is reduced \refs\CDGR.

\item{2.} $B$ field has one leg along the brane and the other leg
perpendicular to it. This is the {\bf dipole theory}.
The brane is no longer pinned
but now the hypers on the world volume develop a
constant dipole length. The vectors
still have zero dipole length \refs{\bergan,\berraj,\berdas,\dasjab}.

\item{3.} $B$ field has both the legs on the brane. This is the
{\bf noncommutative geometry} \refs{\DH, \nekra, \seiwit}.
The theory also develop dipoles but the
dipole lengths are not constant rather they depend on the
momentum \refs{\jabdip, \ZYIN}.

\item{4.} $B$ fields have different (but constant) values on
parallel branes. This is the {\bf nonabelian geometry}.
The theory mixes both the
noncommutativity and the nonabelian nature in a single star
product \refs\KesYin.

\noindent All these theories have been studied earlier. Some in great details and some
in not so great details. However all these theories, though look
different, do actually arise from a single configuration in M-theory which
we shall elaborate in details here. We shall review some of these earlier
works, now in the light of the unified picture,
and provide many new examples of these theories. The M-theory framework
which unifies the set of theory described above is simply a background
with a M5 brane probing a Taub-NUT with G-fluxes. We shall discuss
this in section 2.

But there is more. The four dimensional
Taub-NUT background (which is probed by branes) could
be fibered non-trivially over a $P^1$ base. In fact this background is
related to a six dimensional conifold type geometry with fluxes! The
M5 probing this background has now new set of theories which we shall
elaborate in section 3.

We could even extend this further by making the Taub-NUT geometry multi-
centered. In the presence of various different choices of {\it localised}
G-fluxes, the nonabelian geometry could be described. We will show in section 4
how various properties of these geometries arise from a multi Taub-NUT
background.

In section 5 some applications of the various scenarios will be presented.
We will also point out some possibilities of getting new theories that
are somewhat orthogonal (though related) to the material presented in the
earlier sections. In fact we will argue that the multi Taub-NUT geometry
studied in section 4 can now be fibered over a $P^1$ (as in section 2) to get
a multi-conifold like geometry; and therefore will give a set of new theories.
We will present our conclusions in section 6 and show that even time dependent
backgrounds can be tackled using the generic scheme presented in the
earlier sections.

\section{Theory Near Single-Centered Taub-NUT Geometry}

What we require is the metric of a Taub-NUT space in M-theory and
in the presence of G-fluxes when we allow maximal possible rank
for the fluxes. The metric is in fact been worked out in details
in \refs\cita\ and therefore we shall be brief. The ranks of the
fluxes can actually be incorporated in the background metric by
choosing different harmonic functions along various directions in
M-theory. If we denote the harmonic functions by $H_i$, with $i =
0, 1, 2, 3$, then the metric of the Taub-NUT space modifies from
the usual form to \eqn\tanmetmod{ds^2_{TN} = f^{1\over 3} (dr^2 +
r^2 d\Omega^2) + H_0 f^{-{2 \over 3}} \left(dx^7 + {1\over
2}R~{\rm cos}~\gamma d\phi + ...\right)^2 \, ,} where
$(r,\gamma,\phi)\leftrightarrow  x^{8,9,10}$ and the dotted terms
denote the mixing between $dx^7$ and $dx^0$ as explained in
\refs\cita. We have also defined $f$ as $f = H_1 H_2 H_3$. All the
harmonic functions $H_i$ are functions of $r, R$ and ${\rm
cos}~\alpha_k$ with $k = 1, 2, 3$ and $R$ is related to the radius
of the TN space. The angles $\alpha_k$ are related to the twists
of the T-dual tori as shown in \refs\cita. Now a single centered
TN space supports a unique harmonic form $\omega$, which, in the
presence of fluxes can be written in terms of the above set of
coordinates as $d\zeta$, where \eqn\harmofo{\zeta = g(r)\left(
dx^7 + {1\over 2}R~{\rm cos}~\gamma~d\phi \right),} with $\omega$
satisfying a normalisability condition such that it is anti
self-dual. For the most generic case when we have at most four
different warp factors denoted by $H_{0, 1, 2, 3}$ related to (as
we shall see later on in this section) different ranks of the $B$
fields, we can write the function $g(r)$ as (in units of $2\pi
l_p^2$) \eqn\funcgras{g(r) = {\rm exp}~\left[-{1\over
2}R~\int_r^{\infty} H_0^{1\over 2}H_1^{-{1\over 2}}H_2^{-{1\over
2}}H_3^{-{1\over 2}} \right]\, .} It turns out that for this
generic case a closed form estimate is rather difficult. However
we can try the rank four case. The two form can now be computed
exactly, the function $g(r)$ being given by (again in units of
$2\pi l_p^2$) : \eqn\grfunct{g(r) = \left( 1 + {{\rm cos}~\alpha_2
\over {\rm cos}~\alpha_1} \right) \left(1+ {{\rm cos}~\alpha_1
\over {\rm cos}~\alpha_2}\right) {2r \over \left( \sqrt{R +
{2r~{\rm cos}~\alpha_1 \over {\rm cos}~\alpha_2}} + \sqrt{R +
{2r~{\rm cos}~\alpha_2 \over {\rm cos}~\alpha_1}} \right)^2}\, ,}
where $R$ and $r$ are defined above. There are two cases now:

\item{1.}  $\alpha_1 = \alpha ,\quad  \alpha_2 =0$. This is the
case that has been studied earlier and is used to get
non-commutative theory on D6 branes \refs\cita. We will discuss
this soon.

\item{2.} $\alpha_1 = \alpha_2 = \alpha$. In this case $g(r)$ is
given by $g(r) = 8r(R + 2r)^{-1}$, and is independent of the angle $\alpha$.
However this value of $g(r)$ only gives approximate
results for the noncommutative case. This will be elaborated later.

\noindent In this section we will see how, many interesting dynamics of oriented $B$
fields can arise directly from the TN configuration with the simplest choice of
$g(r)$. We begin with the pinned brane scenario where we will also have a
M5 brane in the TN configuration.

\subsection{Pinned Branes}

The Taub-NUT space is along directions ($x^7, x^8, x^9, x^{10}$), where
$x^7$ is compact circle whose radius $R$ shrinks to zero at the origin
$r = 0$. An M5 brane is kept at
$r=0$ orthogonal to the TN space and is oriented along ($x^0, x^1,...,x^5$).
We identify $x^1$ to the M-theory circle. Now we switch on a 3-form
background $C_{167}$ which is a constant at infinity and is a function of
$r$ given by (in units of $M_p^3$)
\eqn\tnmetric{ C_3 = {C(1+C^2)\over C^2+\left(1+{R\over
r}\right)}~ dx^1\wedge dx^6\wedge (dx^7+A_idx^i)\, ,}
where C is the expectation value of the 3-form at infinity and
$A_i$ are the Taub-NUT gauge fields. It turns out that if the M5
brane has a tension $T_0$ at infinity, then the tension at the
origin of the TN is $T = T_0(1+C^2)^{-{1\over 2}}$. Therefore the
brane is pinned at the origin.

This pinning of the brane gave mass to four scalars of the ($2,0$) theory.
The fermion mass puzzle was solved by observing that the theory has
only a ($4+1$)- dimensional poincare invariance and therefore only ($4+1$)-d
supersymmetry. In ($4+1$)- d a fermion can be given a mass.

Another apparent puzzle arises when we try to study the near
horizon geometry of the M5 brane. According to \refs\CDGR, near
$r=0$ we can scale the coordinate $x^1\to x^1~\sqrt{1+C^2}$ such
that Lorentz invariance is seemingly restored. One might expect
that the near horizon geometry will be  $AdS_7$. This would also
imply that the theory is scale invariant and the mass of the hyper
is zero. But a calculation for fermions show that they are not
massless. The resolution is that the scaling of coordinates which
generated Lorentz invariance makes no sense for a point
arbitrarily close to $r=0$. A small fluctuation of the M5 brane
will break Lorentz invariance in this picture and also conformal
invariance generating a scale $m$ such that \eqn\mdef{ m = {C\over
\sqrt{1+C^2}}\quad .} Therefore, for any finite $C$ the theory no
longer has any $AdS$ limit.

When the external parameters were carefully chosen this lead to
new theories in six dimensions which were decoupled from gravity.
The first decoupled theory can be seen when the external
parameters are: \eqn\decoup{ C \to \epsilon,~~~ R_7 \to
\epsilon,~~~ M_p \to \epsilon^{-\beta},~~~ \beta >1 \quad . } In
this limit the energy scale of the excitations of the M5 brane is
kept finite whereas the other scales in the problem are set to
infinity. This decoupling is kinematical. For a different scaling
of external parameters \eqn\lmiti{ C \to {\rm finite},~~~ R_7 \to
{\rm finite},~~~ M_p \to \infty,~~~g_s \to 0} we get a dynamical
decoupling. This decoupling is in the same spirit as the little
string theory.

\subsection{Dipole Theories}

To get dipole theory from the configuration studied above is easy.
Instead of orienting the M5 along $x^{0,1,2,3,4,5}$, we now orient
the M5 along $x^{0,1,2,3,4,6}$ such that the background $C$-field
$C_{167}$ will have two of its components along the M5 brane. In
type IIA side we have a D4 brane along $x^{0,2,3,4,6}$ with a $B$
field $B_{16}$ along the brane. This case have been studied in
much detail in \refs{\bergan,\berraj, \berdas, \dasjab}\ and
therefore we shall be brief and only mention some of the salient
features. A D4 brane oriented along $x^0, x^2,...,x^5$ and
orthogonal to a Taub-NUT space along $x^7,..., x^{10}$ is pinned
by a pinning potential given by \eqn\pinpot{ {\sqrt{{\rm
det}~g}\over g_s} = {\rm cos}~ \theta = {1\over \sqrt{1+b^2}}\, ,}
in the presence of a $B_{NS}$ field, with an asymptotic value of
$b$, and oriented along $x^6,x^7$ directions as \eqn\bvalve{
B_{NS} = h {\sqrt f_2}~{\rm tan}~\theta~dx^6 \wedge (dx^7 + B_{7i}
dx^i)} where the string coupling $g_s
 = e^{\phi} = \sqrt{h~f_1~f_2^{-1}}$ and $f_{1,2}$
are the harmonic functions for this background with metric
components denoted by $g_{ij}$ and string coupling $g_s$. The
$B_{NS}$ field asymptotes to ${\rm tan}~\theta \equiv b$. In the
above calculations we have defined a quantity $h$ as
\eqn\defaquan{ f_2^{1/2}~h^{-1}= f_2~{\rm sin}^2~\theta + f_1~{\rm
cos}^2~\theta \, ,} which determines how the Taub-NUT circle
behaves in the presence of the $B_{NS}$ field. As its known, the
Taub-NUT circle is non trivially fibered over the base and the
metric of the fibration involves $h$ linearly. When we make a
T-duality along $x^4$ the dilaton changes to $\phi \to \phi -
{1\over 2}{\rm log}~ g_{44}$ and therefore a D3 brane is also
pinned when $B_{NS}$ field is orthogonal to it. The pinning
potential is the same.

This part of the story is well known and therefore we should ask what
happens for the other case, i.e. when we have a D4 oriented along
$x^{0,2,3,4,6}$ at the $r = 0$ point of TN space along $x^{7,8,9,10}$ and
a $B_{67}$ field on it. As shown in
\refs\berraj\ there is no pinning potential
now, but the theory is governed by a new star product called the dipole
star product. Many new aspects of this theory have been studied in
\refs\difuli\ using supergravity analysis and in \refs\difnew\ from the
pp-wave analysis.

An alternative way to get dipole theories on branes is to use duality twist
discussed in \refs\berdas. Equivalence between this and the one got from the
Taub-NUT example is shown in \refs\dasjab.
Furthermore there are two interesting
aspects of these theories not so intuitive:

\item{1.}  There exist a scale $-$ the dipole length $-$ in these theories.
Yet the Beta function is zero. This is also reminiscent for the case of
noncommutative geometry.

\item{2.}  The theory is non-renormalisable in the usual sense because
there are in principle, infinte counterterms. However the form of the
counterterms are exactly determinable at all loop orders.

\subsection{Noncommutative Geometry}

However in both the cases above we have kept the value of $C$ very low.
An interesting case is when $C\to \infty$ and we remove the M5 brane
{}from the picture. We also identify the M-theory cycle as $x^7$ instead of
$x^1$.
In type IIA we will therefore have a D6 brane oriented
along $x^{0,1,2,3,4,5,6}$ with a $B$ field.

Let us first consider the rank two case. We have the $B$ field
oriented along $x_{1,6}$ which we denote by $B_{16}$. When we lift
this to M-theory we have a threeform $C_{167} \equiv C$ where
$x^7$ is the 11th direction as mentioned above. This threeform
will backreact on the geometry and will change the value of $g(r)$
as calculated for the generic case in \grfunct. For our case the
value of $g(r)$ have been worked out earlier in \refs\cita\ and
can also be derived directly from \grfunct\ assuming $\alpha_1 =
\alpha, \alpha_2 = 0$. It is given by \eqn\grforthis{g(r) =
{2r~(1+ {\rm cos}~\alpha)(1 + {\rm sec}~\alpha) \over \left(
\sqrt{R+2r~{\rm cos}~\alpha} + \sqrt{R + 2r~{\rm sec}~\alpha}
\right)^2}.} Using this value of $g(r)$ and considering the
following limits of the background fields (see also \refs\shahin):
\eqn\suly{ C \to \infty,~~~M_p \to \infty,~~~M_p^3 C^{-1} \to {\rm
fixed}} we get a $6+1$ dimensional noncommutative YM theory whose
coupling constant \eqn\ymcoup{ g^2_{YM} = M_p^{-3}C = {\rm
fixed}\, .} This limit is consistent with (and in fact it's the
same as) the limit studied by Seiberg-Witten \refs\seiwit. A way
to see this would be as follows:

The SW limit for a D6 brane with a rank 2 {\it dimensionless}
$B_{\mu\nu}$ field is given in terms of {\it dimensionful} type
IIA metric $g_{\mu\nu}$. If $l_s$ is the string length and $g_s$
is the string coupling constant the limit is \eqn\limiis{ l_s \to
\epsilon^{1/4},~~~g_{\mu\nu} \to \epsilon,~~~ g_s \to
\epsilon^{-1/4} \, ,} where $\epsilon \to 0$ is used to
parametrise very small quantities. We now want to study this limit
{}from M-theory point of view. Recall that we have identified the
eleven dimensional circle with the Taub-NUT circle, i.e with $y$
and taken the length dimensions in M-theory as \eqn\lengthdim{
[~g^M_{mn}~] = 0, ~~~~ [~C_{mnp}~] = 0,~~~~ [~\partial~] = -1 \,
,} where $g^M_{mn}$ is the 11d metric. Invoking now the usual
relations between the IIA variables and M-theory variables, we are
led to the following limit: \eqn\follim{ R_y~ \simeq ~g_s^{2/3}
l_p ~\simeq ~g_s l_s = {\rm constant}, ~~~~l_p \to \epsilon^{1/6}
\, ,} where $R_y$ is the radius of the Taub-NUT circle. This would
imply that in our limit when
 $l_p^3 C$ is a constant,
\eqn\cis{ C \to \epsilon^{-1/2}} and the type IIA variables
$B_{\mu\nu}$ and $g_{\mu\nu}$ are related to the dimensionless $C$
field as $C = \frac{\alpha' B}{g}$. A consistency check of this
would be to note that \eqn\conscheck{ g^2_{YM} =
g_s~l_s^3~\sqrt{{\rm det} \left( \frac{\alpha'B}{g} \right)} \to
{\rm constant} } agrees with the identification. Therefore to
summarise, we get a $6+1$ dimensional noncommutative YM for the
following limit of the external parameters: \eqn\extpara{ C \to
\epsilon^{-1/2},~~~~ M_p \to \epsilon^{-1/6},~~~~ R_y \to {\rm
constant}\, .} Observe that the $C$ field in other directions are
not excited. The dimensionless M-theory metric, $g^M_{mn}$, now
scales differently. Along the directions of non-commutativity they
scale as \eqn\scaleas{ g^M_{55} \to \epsilon^{2/3}, ~~~~g^M_{66}
\to \epsilon^{2/3}\, . } Along other directions they scale in the
following way: \eqn\scnow{ g^M_{\mu\nu} \to \epsilon^{1/6}, ~~~~
g^M_{yy} \to \epsilon^{-1/3} \, . } It is easy to check that
$R^2_y = l_p^2 g^M_{yy} \to {\rm constant}$ as expected.

\vskip 5pt

\noindent~1.~$\underline{\rm The~rank~four~and~rank~six~cases}$

\vskip 5pt

To study the rank $r$ case, we will use the fact that string
coupling scales as \eqn\ranf{ g_s \to \epsilon^{r-3\over 4}, ~~~~
{\rm and}~~~~ l_s \to \epsilon^{1/4}\, . } An obvious invariant
that could be made from above scaling is the combination: $g_s
l_s^{3-r}$. Indeed this is exactly what is kept fixed in the
non-commutative theory and is related to the $g^2_{YM}$. In the
M-theory description we have \eqn\mthe{ R_y \to \epsilon^{r-2\over
4}, ~~~~{\rm and}~~~~ l_p \to \epsilon^{r/12}\, .  } This
immediately implies that for the higher rank fields we really do
not need the 11-dimensional picture at the decoupling limit. This
is also consistent with earlier results \refs\alisha. We notice
also that the 11-dimensional Planck length does go to zero for
rank 4 and 6 cases too. The YM coupling constant is given in terms
of M-theory variables as: $l_p^3 C^{r/2}$. This implies
\eqn\cisno{ C_i \to \epsilon^{-1/2}, ~~~~{\rm and} ~~~~
 \sqrt{{\rm det}\left({\alpha'B\over g}\right)}
\to \epsilon^{-r/4}\, ,} where $i=2,..,r$ and therefore $g^2_{YM}$
is a constant. The metric along the non-commutativity directions
now scale as \eqn\comdir{ g^M_{ij} \to \epsilon^{6-r\over 6} } and
the dimensionless and the dimensionful metrics in the 11th
direction $y$ scale as \eqn\elevdir{ g^M_{yy} \to \epsilon^{r-3
\over 3}, ~~~~ g_{yy} = l_p^2 g^M_{yy} \equiv R_y^2 \to
\epsilon^{r-2\over 4} \, . } Along other directions they scale as
\eqn\sclll{ g^M_{mn} \to \epsilon^{3-r\over 6}\, ,} implying that
the metric scales differently along different directions of the
brane as expected. Similar behaviour can be seen for the dipole
theory also which was mentioned in \refs\berraj.

\vskip 5pt

\noindent~2.~$\underline{l_p~ \to~ 0~{\rm limit}}$

\vskip 5pt

In the M-theory description we define various scales as
\eqn\varsca{ 2\kappa_{11}^2 = (2\pi)^5 l_p^9, ~~~\kappa_{10}=
{8\pi^4 l_p^9 \over R_y}, ~~~ T= {1\over 2\pi l_p^3} \, ,} where
$T$ is the membrane tension. The coefficient of the bosonic part
of the 11d supergravity is $-{1\over 2\kappa_{11}^2}$ if we take
the following length dimensions \eqn\follen{
[~g~]=0,~~~~[~C_i~]=0,~~~~[~\partial~]=-1 \, . } To see whether
the gauge kinetic term survives let us consider the kinetic term
of the three-form field. It is given by \eqn\givby{ {1\over
2\kappa_{11}^2} \int ~G \wedge \ast_{11} G ~\to~ l_p^{-9}\int~
\sqrt{g}~G_{\mu\nu\rho\sigma} G_{\mu'\nu'\rho'\sigma'}~
g^{\mu\mu'}g^{\nu\nu'}g^{\rho\rho'}g^{\sigma\sigma'} \, ,} where
$g_{\mu\nu} \equiv g^M_{\mu\nu}$ is the dimensionless M-theory
metric. If we require the gauge fields to be along directions
$x_{1,2,3,4}$ then a typical term will look like \eqn\typsca{
l_p^{-9}\sqrt{g}~g^{11}g^{22}g^{mm'}g^{nn'}~G_{12mn}G_{12m'n'} \,
. } It is easy to see that this term scales as \eqn\terra{
\epsilon^{-1/2}~g^{-1}_s~l_s^{-3} \, ,} which is a constant
implying that the gauge kinetic term survives the scaling.

Consider now the next higher order term \eqn\nexhi{ T\int \sqrt{g}
R^{hmnk}R_{pmnq}R_n^{~~rsp}R^q_{~~rsk} ~~\sim ~~ l_p^{-3}
\sqrt{g}(g^{-1})^8 R^4 \, . } Each $R$ scales as either
$\epsilon^{2/3}, \epsilon^{1/6}$ or $\epsilon^{-1/3}$ depending on
the orientation. In fact it is easy to see that \eqn\orient{
\epsilon^{1/6}~ \le ~ (g^{-1})^8 ~ \le ~\epsilon^{-23/6}, ~~~~
\epsilon^{8/3}~ \le ~ (R)^4 ~ \le ~\epsilon^{-4/3} \, . } This
term decouples in the path integral and therefore do not
contribute to the quantum fluctuations. It would be interesting to
study the behaviour of other higher order terms. However since the
explicit forms of these terms are not known, it is difficult to
see what happens under the scalings.

For this theory we could calculate various BPS states. The masses of these
states are all proportional to $(M_p^3/C)^{\alpha}$. We get light M2 branes
for $\alpha =1$ and light M5 branes for $\alpha = 2$.

\section{Theory Near Conifold Type Geometry}

In the previous section we studied an M5 brane near a Taub-NUT
singularity. Let us reorient the system such that the M5 is
oriented along, say, $x^{0, 1, 2, 3, 4, 9}$ and the TN along
$x^{6, 7, 8, 10}$ with $x^7$ being the TN circle and the other
three directions non-compact. Furthermore the directions $x^{5,9}$
are also non-compact. Assuming now that the directions $x^{5,9}$
have a topology of a sphere $P^1$ the ALE space will have the
following form: \eqn\alefib{ z_1^2 + z_2^2 + z_3^2 = - \vert \mu
\vert^2 \, ,} where $z_{1,2,3}$ are used to denote the TN space
and $\mu$ is the size of the blown up sphere $P^1$. If we now
identify $\mu = z_4 = x^5 + i x^9$ then this is the equation of a
conifold oriented along $x^{5,6,7,8,9,10}$. Thus we get a way to
study branes near conifold singularities using the approach
discussed in the previous section!

To be a little more precise, we actually need a configuration of M5 brane
near a conifold type singularity in the presence of fluxes. To achieve the
supergravity description of this we take
a system of intersecting D5 and NS5 branes in type IIB theory. The
D5 brane is oriented along ($x^0, x^2,...,x^4, x^7, x^9$) and we have two NS5
branes oriented along ($x^0, x^2,....,x^6$) and ($x^0,...,x^4, x^8, x^9$)
respectively. We will further assume that directions $x^6, x^7$ are on a
slanted torus whose angle is $\theta$. The direction $x^1$ will be the
M-theory direction. By T-dualising along $x^7$ we get a configuration in
type IIA theory\foot{A slight variant of this problem was studied earlier
in \refs{\angle,\DM} where a conifold in type IIB theory was shown to be
T-dual to two intersecting NS5 branes. Another related construction was
given in \sadovafa\ where a conifold in type IIB theory was shown to be
T-dual to two intersecting NS5 in type IIB theory giving rise to the
so-called {\it brane boxes} \refs{\karch}. These construction (in the
IIA picture) were used, in a different context, to understand the
geometric transition in ${\cal N} = 1$ theories \refs\geometric. Although
unrelated to the main line of thought presented here, the Taub-NUT background
also played a very crucial role there. We will however not discuss this
anymore in the paper.},
which when lifted to M-theory along $x^1$ will reproduce the
background that we are interested in.

\subsection{Pinned Branes}

The interesting thing now is that the harmonic functions are all
{\it linear} functions of the overall transverse direction, which is
$x^{10}$ in our case. This is however assuming that the D5 brane is
completely delocalised along the NS5 branes' directions.

We have oriented the D5 brane in such a way that the theory on it
will be a $U(1)\times U(1)$ gauge theory. The product gauge group
arises simply because the D5 brane is ``cut'' twice by the two NS5
branes. The theory will now have a lower supersymmetry. The metric
for this configuration will be same as of D5-NS5-NS5' branes but
with a slight deformation along the $x^6, x^7$ directions because
of the slanted torus. There will be a cross term in the metric
which as seen by the D5 brane will be \eqn\coni{ 2 (1+\vert
x^{10}\vert)~~ {\rm tan}~\theta ~~dx^6 dx^7 \, . } The reason we
choose this configuration is because by making a T-duality along
direction $x^7$ we will get a configuration of a D4 brane at a
conifold point in type IIA. The conifold in question arises from
the two intersecting NS5 branes which overlap along three common
directions. We will also get a NSNS $B$ field background from the
slant of the torus. Note that the two NS5 branes also contribute
to NS $B$ fields. We can gauge away most of
 the components of the source $B$ fields and keep only $B_{78}$ and
$B_{75}$ from the two NS5 branes respectively. Therefore the
background now looks like \eqn\conitwo{ B = {{\rm tan}~ \theta
\over {\rm sin}^2\theta + (1+\vert x^{10}\vert)^q~{\rm
cos}^2\theta}~~ dx^6\wedge (dx^7 + B_{78}dx^8 + B_{75}dx^5)\, ,}
where $q=1$ when we have the D4 as a probe, and $q=0$ when the
effect of D4 is completely delocalised.

It is important to note that we actually get a conifold for values of
$\vert x^{10}\vert$
sufficiently small. For this limit all the harmonic functions
are essentially constant and the metric looks conformal to a conifold
under some scaling. Another crucial thing which is necessary to reproduce
the conifold geometry is that directions $x^{4,5}$ and $x^{8,9}$ are
spheres. However for our case we will not take these directions as spheres,
instead they will be toroidal. Therefore the geometry is conifold like
(meaning that it falls in the same equivalent class as conifold geometry).

We now lift this configuration to M-theory. The parameter which lifts the
metric is the type IIA coupling given by
\eqn\cone{
e^{2\phi} = {(1+\vert x^{10}\vert)^p  \over A~{\rm sin}^2\theta +
B~(1+\vert x^{10}\vert )^q~{\rm cos}^2\theta}.}
There are three interesting cases now:

\item{1.} In the absence of probe branes, $A = B = p = q = 1$. This is the
usual case that we shall be concentrating mostly. This behaviour persists
when we take the effect of probe on the background to be very small.

\item{2.} When we assume full delocalisation in the presence of probe
 branes $A = B = 1, q = 0, p = -{3\over 2}$. This case is also
similar to the case when we have $q = 0$ in \conitwo. The delocalisation
effect dilutes the background considerably and therefore most of the
effects are actually washed out, for example the non-trivial behaviour of
$B$-field.

\item{3.} Assuming partial delocalisation, $p = q = 1$ and $A = f^{5\over
2}, B = f^{3\over 2}$; where $f$ is the localised harmonic function for the
probe. This is a more involved case and we will have nothing new to say here.

\noindent In the absence of any background B-field $e^{2\phi} =1$ and therefore the
M-theory and the type IIA metric is the same. For this case we see that
we can come back to type IIA via the $x^7$ circle (instead of $x^1$). But now
we will get a configuration of two intersecting D6 branes (see also
\refs\tataroh). This
information can also be used to verify the results.

Another important thing to note is that the M5 brane (which is the lift of
the D4 brane) will only see a $3+1$ dimensional poincare invariance. This is
 because of our choice of the two intersecting NS5 branes. The M5 brane
overlaps with four of the ``flat'' directions of the conifold but one of
the directions ($x^1$) is scaled differently.

If $T_0$ is the tension of the M5 brane in the absence of the background
$B$ field then near the origin $\vert x^{10}\vert  \to 0$
we have a M5 brane with
tension $T_0$ and near  $\vert x^{10}\vert  \to \infty$ we have a M5 brane with
tension $T_0~{\rm cos}\theta$.

Let us consider a case in which the slant of the torus $\theta \to
\pi/2$. In this situation, writing the metric components as
$g_{ij}$ before, we observe that for all finite values of $\vert
x^{10}\vert$ \eqn\metty{ det~g =
~g_{00}~g_{11}~g_{22}~g_{33}~g_{44}~g_{99} = ~ {\rm sin}^2 \theta
+ (1+\vert x^{10}\vert) {\rm cos}^2 \theta \to 1 \, . } This
result seems to suggest that a M5 brane will now see a completely
flat potential! Therefore any pinning in the {\it absence} of $C$
is completely removed in this limit. However the way we have
motivated the model - starting with D5 -NS5 -NS5' system - the D5
is stuck at the zero of the coulomb branch by the construction
itself. There is of course the Higgs branch but that motion will
give rise to some massive states in the gauge theory. Observe that
for the case of M5 brane near a Taub-NUT space, even for $C\to
\infty$, this effect will never be there. In fact we will get the
maximum pinning for this limit. The tension of the M5 brane will
remain 1 at $r\to \infty$.

\noindent Thus we seem to be getting the
following salient features from our model:

\item{1.} Theory on the M5 brane
(or D4 brane from IIA point) has only $3 +1$
dimensional poincare invariance. The $4+1$d poincare invariance here is
broken by the geometry of the construction, as against the model studied
earlier, where the poincare invariance was broken by the background $C$
field.

\item{2.} From construction we
have a M5 brane ``pinned'' at the origin. But
now, for the choice of large background or $\theta \to \pi/2$, we can keep
another M5 brane {\it anywhere} in the spacetime, which would not be possible
for $\theta = 0$. This effect is completely opposite of the effect we
saw in the previous case. Here we are getting an unpinning of the brane
because of the presence of large $C$ field.
 This may be used to study models analogous to the
the Brane-world scenario  where the starting point is by keeping
a mirror brane at a distance $r_c$ from the original brane and then take the
limit of $r_c \to \infty$.

\item{3.} The theory on the M5 brane
now has only $N =1$ supersymmetry in the
poincare invariant $3+1$ dimensional space.

\noindent One can continue along these lines to study more interesting aspects of
pinned (or ``unpinned'') branes. As such it is not clear whether the pinning
effect remains when we have a configuration which breaks supersymmetry
completely. For the conifold like singularity we do not have an identical
BPS calculation, which was done for the Taub-NUT case, to support our
observation of unpinning. It will be interesting to find an alternate
confirmation of this.

\subsection{Hybrid Theory: Noncommutative Geometry and Dipole Theory}

In the above analysis we briefly mentioned that we could get
intersecting six-branes in type IIA theory if we reduce the
conifold geometry along some other direction. Let us recapitulate
the issue. We started with a configuration of a D3 brane near two
intersecting NS5 branes. The metric of the system is more or less
identical to the metric of a single D3 brane without any other
effects. The back-reaction of the NS5 branes can be incorporated
by shifting the coordinates $dx^i$ to \eqn\shiftdx{dx^i~ \to ~ (1
+ \vert x^{10} \vert)^{p/2}~dx^i,} where $p$ can be either 0, 1 or
2 depending on the directions. As discussed earlier, the
background also have $B$ fields that are sources of the NS5
branes. For simplicity if we assume that these backgrounds are
unity for some appropriate choice of scales, then it is easy to
show that the conifold cycle in M-theory (which we denote by
$d\psi$) becomes \eqn\conicyclebe{ d\psi = {dx^7 +{\rm cos}
\theta_1~dx^8 + {\rm cos} \theta_2~dx^5 \over (1 + \vert x^{10}
\vert)^{2\over 3}~(1+\vert x^{10} \vert {\rm cos}^2 \theta)^{1
\over 3}},} where ${\rm tan}~\theta$ measures the value of the
antisymmetric field at infinity and $\theta_{1,2}$ measure the
values of B-fields $B_{78}$ and $B_{75}$. The above way of writing
also guarantees that the M-theory background can be put in a
simple form as \eqn\mthmet{ds^2 = d\psi^2 + e^{4\phi/3}~(dx^1)^2 +
e^{-2\phi/3}~[-(dx^0)^2 + (dx^2)^2 + ...] \, ,} where we have used
the scalings given in \shiftdx. There is a small subtlety though.
In the presence of G-fluxes, the direction $dx^6$ is further
suppressed by ${\rm sec}~\theta$, even though rest of the
components follow the rule given in \shiftdx.

At this point we have two possibilities to come down to type IIA theory: either
along $x^1$ or along $x^7$. The first possibility is of course the one dealt
in the previous section, wherein we get conifold singularity in the type IIA
side. The novel thing is the other way of doing it. It is well known that
in the absence of any fluxes, reducing M-theory on $x^7$, will give us two
intersecting D6 branes \refs\tataroh.
What happens in the presence of fluxes? We will indeed
get two intersecting D6 branes, but now due to the presence of fluxes, one of
the branes will have non-commutaive star product whereas the other one will
have dipole star product! This is what we call as {\it Hybrid Theory}.
Question is how is this consistent, when, on one side spacetime coordinates
do-not commute whereas the other side they do?

To answer this, observe that we get two D6 branes oriented along
$x^{0,1,2,3,4,5,6}$ and $x^{0,1,2,3,4,8,9}$ with a curved metric that can be
easily determined from the above analysis\foot{For the case when we have
two NS5 branes oriented along $x^{0,2,3,4,5,6}$ and $x^{0,2,3,4,8,9}$, an
S-duality transformation will give us two intersecting D5. Now T-dualising
along $x^7$ we will get two intersecting D6 branes in type IIA theory.
Lifting this to M-theory along the M-theory direction $x^1$,
this will give a conifold like geometry. The relation between the intersecting
D6 that we have here and the one got by S-duality of the NS5 is simply the
interchange of $x^1, x^7$ direction in M-theory. As is well known, this is
how the S-duality shows up in M-theory.}.
There would be the KK gauge fields
that would come directly from the cross terms discussed in the metric
component $d\psi$. These KK gauge fields are of course the D6 brane sources.
There would also be RR three forms $C_{168}$ and $C_{165}$ plus an
antisymmetric two form
\eqn\bford{ B_{16} = {{\rm tan}~\theta \over 1 +
\vert x^{10} \vert~{\rm cos^2} \theta}.}
This is the two-form that is responsible for generating a hybrid theory.
Observe that now, even though we do not have any TN background, this $B$
field cannot be gauged away because it lies on both the branes. Because of the
orientation of the two D6 branes there is no conflict due to spatial
non-commutativity as the directions of non-commutativity do-not overlap
completely. In other words, the direction $x^1$ and $x^6$ are non-commutative
but $x^6$ do not lie on the other brane. Since $x^1$ commutes with
$x^{2,3,4,8,9}$ directions on the other D6 brane, it doesn't violate the
fact that this theory can now have dipole star product.

\section{Theory Near Multi-Centered Taub-NUT Geometry}

Till now we have either taken a single centered Taub-NUT space or a Taub-NUT
space fibered over a $P^1$ base forming a conifold. It is time now to go
to multi-centered TN space in the presence of fluxes and see whether we can
extend the previous analysis to this level also. We could take, as before,
an M5 brane near such singularities (in the presence of oriented fluxes) or
remove the M5 brane and study the configuration directly in type IIA picture.
Let us first consider the case when we have no M5 branes. Let us also assume
that all the TN singularites are at the point $r = 0$.

In type IIA theory this therefore gives rise to multi D6 branes stacked
at the point $r = 0$.
In the presence of large number of D6 branes there is an inherent
non-commutativity of the space-time coordinates because of the non-abelian
nature. This is reflected from the fact that matrices don't commute. Let us
now
switch on a $B$ field to generate {\it further} noncommutativity.
However these two noncommutativity do not mix
in any usual sense and the product rule is specified by a simple
tensoring of constant matrix algebra and the Moyal-Weyl deformation
\eqn\moyal{
(\Phi*\Psi)^i_k (X) = \sum_j~ \left(\Phi^i_j * \Psi^j_k \right) (X).}
This is however not quite the case when we have multiple D-branes each
seeing a different B-field on its world volume\refs{\KesYin}.
A way to configure such
a system in string theory is to have multiple D-branes in a spatially
varying B-field\foot{Another way would be to switch on different gauge
fields $F_i$ on the $i^{th}$ brane. In the presence of both $F$ and $B$
the invariant quantity is $\cF_i=F_i-B_i$. Henceforth we will just specify
$\cF_i$.}.
The pull back of the B-field on each brane is constant.
In the setup that we described earlier, this could be realised by
switching on a $G$-flux in the
multi Taub-NUT background. The $G$-flux has
non-zero expectation values only near the Taub-NUT singularities
\refs{\KesYin}.
This configuration
may or may not preserve any supersymmetry depending on the configuration.
Later on we shall give a concrete example where an ${\cal N} = 1$
susy is preserved.

As it turns out, a two point lattice approximation to open string
is perfectly suited for this \refs{\ZYIN}. We can get away with
the enormously complex nature of products of string wave functions
while retaining the essential nature of the non-commutativity of
such products. Therefore, though the calculations are motivated
{}from string theory, the lattice string quantum mechanics (LSQM) is
an independent way to calculate this product. More clearly, what
we need is the following decomposition rule for a string
wavefunction $\Psi(x)$ to the dipole basis $e_{1,2}$
\eqn\wavtodipbase{\Psi(e^a_1, e^a_2) = \int~dX'~\Psi(X')~\langle
X' \vert e^a_1, e^a_2\rangle \, ,} where $\langle X' \vert e^a_1,
e^a_2\rangle$ is the change of basis function that we shall
discuss in the next few sections. A way to evaluate this is given
in \refs{\KesYin, \ZYIN}. Using this one can show that the  new
product is no longer a  simple tensoring of the star product
\moyal\ and non-abelian matrix algebra. The noncommutative {\it
real} space and the non-Abelian internal space get intertwined and
inseparable. This is the main idea of {\it nonabelian geometry}
\refs{\KesYin}. The deformation equation becomes:
\eqn\nonabelian{ (\Phi\times \Psi)^i_k (X)\equiv \sum_j~ \left(
\Phi^i_j *_{ijk} \Psi^j_k \right) (X)\, ,}
where $*_{ijk}$ depends on all the three indices.
Let us now discuss the various possibilities.

\subsection{Pinned Branes And Dipole Theory}

What we now require is that a M5 brane should probe this
background. In IIA theory this is nothing but a D4 probing a multi
TN background, in the presence of $B$ flux that could either be
completely orthogonal to the brane or should have one leg along
the brane. For the case when we have the $B$ field completely
orthogonal to the brane, it is determined in terms of $\alpha$ as
\eqn\bnow{ B = {1 + a~{\rm cos}^2 \alpha \over 1 + b~{\rm cos}^2
\alpha}~B_o \, ,} where $B_o$ is the $B$ field for the single
centered TN. The factors $a,b$ are  in general functions of the
$r$ and the points where the TN circles would degenerate, the
explicit form of which can be easily determined from \refs\CDGR.
Question now is what would happen to the probe M5 brane as it goes
towards any degenerating points of the multi TN space. One can
argue that at any such points the backreaction due to the harmonic
functions is very large and therefore the M5 brane is effectively
pinned at that point. Therefore the multi TN space has many fixed
points for the M5 brane and once the brane is fixed at one such
fixed point it will not move to the other fixed point. Similar
analysis can be performed for the dipole theory. One can show that
the branes are not pinned at any points on the multi TN space and
therefore the dipole behavior is not quite different from the case
of single centered TN space.

\subsection{Noncommutative Geometry}

To understand the noncommutative theory from the point of view of
M-theory, we can use the earlier configuration but now replace the
background with a multi-TN space, as discussed above. In a
$N$-centered TN space there are accordingly $N$ different
normalisable harmonic forms $\omega_i \equiv d \zeta_i$ satisfying
\eqn\harmsat{\int_{TN} \omega_i \wedge \omega_j = (16 \pi m)^2
\delta_{ij}\, ,} where $m$ measures the periodicity $16 \pi m$, of
the TN circle so that there is no conical singularities at any
point where a circle would degenerate. The generic description in
the presence of fluxes would basically follow the arguments
developed earlier and in \refs\cita. For the rank four case
(calling the harmonic functions for the background as
$H_{0,1,2}$), the gauge coupling constant would be different for
the different directions of the $B$ fields. For gauge fluctuations
completely orthogonal to the noncommutativity direction, an
estimate for the coupling is given by the following integral over
the TN space: \eqn\intovtn{\int_{TN} H_0 H_1^{-{1\over 2}}
H_2^{-{1\over 2}}~\omega \wedge \ast \omega \, ,} where the Hodge
star is over the four dimensions. This integral is approximate
because of the reasons mentioned in \refs\cita\ but give a
rough estimate of the result. For the rank two case this integral
does indeed reproduce the correct coupling as mentioned in
\refs\cita. Now for the gauge fluctuation along the
noncommutativity direction, an estimate for the coupling constant
can now be done by the following integral over the TN background:
\eqn\esnowtn{\int_{TN} H_0^{-1} H_1^{{3\over 2}} H_2^{-{1\over 2}}
~\omega \wedge \ast \omega   \, ,} with a similar estimate for the
other directions. This estimate improves when we consider rank two
case. When we consider the case where all the ranks of the $B$
fields are the same, \esnowtn\ will predict a coupling $l_p^3 C^2$
for $g^2_{YM}$, where $C$ is the expectation value of the
bacground $C_{167}$ field. This is more or less what one would
have expected for such a case, suggesting that the above
integrals, though approximate, are not without merit.

Question now would be how to understand the open string behaviour
directly from the TN geometry. This is easy if we assume that the points
where we have shrunk two cycles, there are also wrapped M2 branes. These
M2 branes will in fact appear on the D6 branes as open strings connecting two
such D6 branes. We can take our {\it open} membrane with a cylindrical
topology and with
a coupling to a generalised three form $C + dB$. The boundary condition can be
written down easily. Now we shrink the torus at two ends so that the
topology is of a sphere as shown below:

\vskip.1in

\centerline{\epsfbox{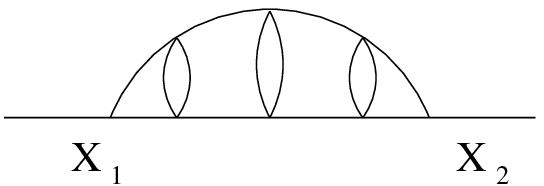}}\nobreak

\vskip.1in

The boundary condition therefore simply becomes the boundary
condition for open strings coupled to some gauge field. This
situation is now ripe for the LSQM approach. The LSQM is the
approximation wherein an open string is represented by two points,
namely the two ends, and labelled by $X_1$ and $X_2$. The open
string action now becomes the motion of a constrained dipole
\refs{\ZYIN}\foot{This dipole nature of the open strings should
not be confused with the dipole theory discussed earlier. The
dipoles of the dipole theory are the behavior of  {\it solitonic}
strings that are constrained to move in the background of $H_{NS}$
field. As a result of this motion, these strings tend to arch out
of the brane on which the two ends are stuck. At a particular
distances between the two ends of the solitonic strings the system
is stable. These two ends form the {\it dipoles} of the dipole
theory \refs\berraj. The dipoles discussed in this section and
next are an approximation where the full string is represented by
its two end points.}. In the zero slope limit $\alpha'\to 0$ while
$\cF$ and $(2\pi \alpha')^2 G^{-1}$ remain finite and $G$ being
the closed string metric, the result of canonical quantization
with constraints is \refs{\KesYin,\SJ}
\eqn\dipolequantiz{\eqalign {
    \comm {X_1^\mu} {X_1^\nu} &=~ \imath \Omega_i^{\mu\nu}, \cr
    \comm {X_2^\mu} {X_2^\nu} &= - \imath \Omega_j^{\mu\nu}, \cr
    \comm {X_1^\mu} {X_2^\nu} &=~ 0.}}
{}From M-theory point of view these commutation relations are the one
for the end points of an open membrane. Since the end points are shrunk to
points, the commutation relations are simpler then their cousins.
Using congruence transformation we can always transform $\Omega$ (or
$\cF$) into the following canonical form:
\eqn\CanonicalFormForF {
    \Omega_i = (\cF^i)^{-1} = T_i J T_i^\top .}
The above way of writing will simplify the formulation of $*$-product in the
subsequent analysis. Also notice from \dipolequantiz\ $X_1$ and $X_2$
commutes among themselves. Actually in $R^D$ there are only $D/2$
commuting coordinates at the two ends. Let us denote them as $e^a_1$ and
$e^a_2$, $a=1,2,..., D/2$, where the subscript indicates boundary
components. If $A,B,...$ denote the rest of the coordinates then the
canonical form $J$ is
\eqn\DefinitionOfe { \eqalign {
    J^{aA} &= \delta^{a+{D\over 2},A} = - J^{Aa} \cr
    J^{ab} &= 0 = J^{AB} .\cr  }}
Therefore the Hilbert space for LSQM is the tensor product of the
Hilbert spaces of the two {\it decoupled} ends of the dipole. In this
formulation, for wave functions $\Psi_{aa}(e_1^a, e_2^a)$
and $\Phi_{aa}(e_1^a, e_2^a)$
in the $aa$ representation, the
defination of $*$-product is straightforward:
\eqn\aaProduct {
    \paren {\Psi * \Phi}_{aa}(e_1^a, e_2^a)
    = \myI {dM^a} \Psi_{aa}(e_1^a, M^a)  \Phi_{aa}(M^a, e_2^a), }
where the equality is upto a normalisation. This way of
formulating the $*$-product, other than its intuitive appeal, has
its root in the string field theory. It basically tells us the
merging of two oriented paths into one as in string field theory
\refs\WITSFT  or, here, as the merging of two ordered pairs of
points. To see that \aaProduct\ reduces to the known formulation
of the deformed product let us consider the case when
\eqn\conscase{ \Omega_i = \Omega_j = \Omega \, .} As discussed
above, since all the branes see the same noncommutativity
parameter $\Omega$, we can convert $\Omega$ to its canonical form
$J$. In this limit, since all $T_i$'s are identity matrices,
\eqn\tidef{ e_i= T^{-1}_i X_1, ~~~~~e_j=T^{-1}_jX_2 }
are just $e_1, e_2$, satisfying the following commutation relations:
\eqn\CanonicalCommutationFore { \eqalign {
    \comm {e_1^\mu} {e_1^\nu} &= J^{\mu\nu}
    = - \comm {e_2^\mu} {e_2^\nu}, \cr
    \comm {e_1^\mu} {e_2^\nu} &= 0 \, .}  }
Now to apply \aaProduct\ we have to define wave functions in terms of
{\it commuting} coordinates. This can be readily shown to be the center
of mass of the dipole
\eqn\comofdip{
X_c= {1\over 2} (X_1+X_2) = {1\over 2} (e_1+e_2).}
We are still a step behind using \aaProduct. What we now need is the change
of basis function. Recall that the number of commuting coordinates are
denoted by $a=1,2,..,D/2$ whereas $A,B,..$ denote the rest of the coordinates.
The change of basis is therefore \refs\ZYIN
\eqn\SimplestCaseOfChangeOfBasi{
    \langle{X_c} \vert {e_1^a,e^a_2}\rangle
    = \delta \left(X_c^a - {1\over 2} (e_1 + e_2)^a  \right)
    \exp {\left( - \frac \imath 4 X_c^A J_{Aa}
        (e_2 - e_1)^a \right)}.}
With this its now straightforward to use \aaProduct. In terms of more
general form $\Omega$ \aaProduct\ is explicitly given by
\eqn\starproduct    {
    (\Psi * \Phi)(X) = \exp {\paren {\frac \imath 2
        \Omega^{\mu\nu} \frac \pa {\pa{X'^{\mu}}}
         \frac \pa {\pa{X''^{\nu}} }}}
    \Psi(X') \Phi(X'')\Bigg\vert^{X'' = X' = X} \, ,}
which is precisely the deformed product appearing in
noncommutative geometry \refs{\CDS,\DH,\seiwit}. The above
analysis can also be presented in the following graphical way:
\vskip.1in

\centerline{\epsfbox{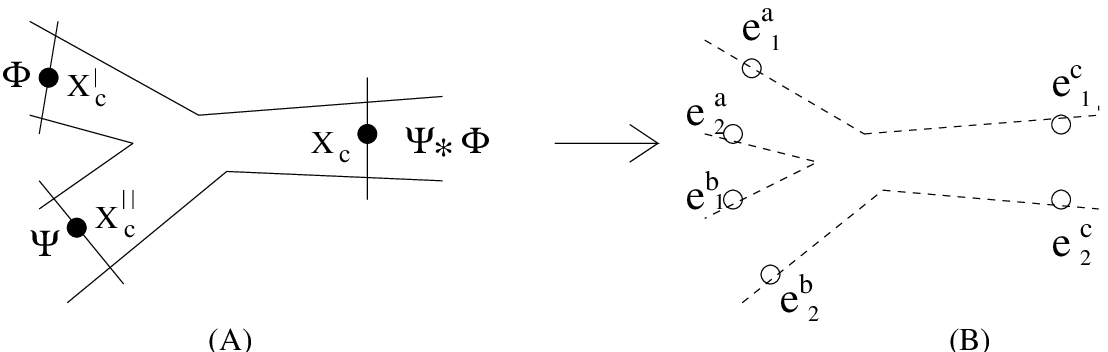}}\nobreak

\vskip.1in
\noindent where figure (A)
represents the way two wavefunctions $\Psi$ and $\Phi$
 merge together to give the product wavefunction $\Psi * \Phi$ and figure
(B) is the corresponding dipole way of
viewing the above process. We have also shown the center
 of masses explicitly. The concept of center of mass will have important
implications when we go to the next section.

\subsection{Nonabelian Geometry}

In the above analysis we
didn't allow the nonabelian nature of the D6 branes to mix with
the non-commutativity by choosing the same  constant $B$ fields on
all the branes. This may not be the case always. The nonabelian
nature of the multiple D6 branes is apparent directly in the
M-theory from the wrapped M2 branes. For this let us assume that
the multi-TN space has many non-vanishing two cycles on which we
can wrap M2 branes. Indeed the intersection numbers of all the
wrapped M2 branes generate the following intersection matrix:
\eqn\intermat{{\cal I} = \pmatrix{2  & -1 & 0  & 0 & \ldots & 0 &
0\cr
           -1 & 2  & -1 & 0 & \ldots & 0 & 0\cr
            0 & -1 & 2  & -1 &\ldots & 0 & 0\cr
            \cdot & \cdot & \cdot & \cdot & \ldots & \cdot &
\cdot \cr
            \cdot & \cdot & \cdot & \cdot & \ldots & \cdot &
\cdot \cr
            0 & 0  & 0  & 0  & \ldots & 2 & -1\cr
            0 & 0 &  0  & 0  & \ldots & -1 & 2}}
as shown by Sen \refs\senm. This intersection matrix is of course
the Cartan matrix for $A_{N-1}$ singularity, which becomes $SU(N)$
gauge symmetry on $N$ D6 branes. Assuming now that we do not
consider the case of coinciding D6 branes (we will eventually
tackle the situation when we have coinciding system), the {\it
nonabelian} geometry is realised now by the following choice of
the background $G$ fluxes: \eqn\gfluxch{{G \over 2\pi} =
\sum_{i=1}^N {\cal F}_i \wedge \omega_i \, ,} where $\omega_i$ are
the harmonic forms satisfying \harmsat. In IIA point of view, this
is then the setup where multiple D6 branes sees different amounts
of noncommutativity on their world volumes. The reader, however,
might still be concerned by the fact that at the coinciding stage
the system might settle down to a configuration with a unique
${\cal F}$ field on its world volume. In the case that we study,
this could in principle happen as a time dependent process but our
main concern is not, for the time being, to study the final stage
of this process. We will take this configuration to illustrate the
possibility of a new star-product and then in the later part of
the paper will give an example wherein such a configuration do
occur. Therefore, in the language of D6 branes and LSQM we will
assume: \eqn\nonanana{ \Omega_i \ne \Omega_j} as our starting
point for nonabelian geometry. Now, as before, to use \aaProduct\
we have to search for the commuting center of mass coordinate.
Here comes the subtlety. The center of mass now is no longer the
simple average of the two ends of the dipole, since such an
average will not commute. However some small algebraic
manipulations will give us the required $X_c$ as
\refs\KesYin\foot{We choose to denote the bases vectors as $e_i,
e_j$ instead of $e_1, e_2$ because of the obvious indications from
\tidef. We hope that this will not confuse the readers.}
\eqn\NormalFormForXij {
     X_c \equiv (T_i^{-1}+ T_j^{-1})^{-1}(e_i +   e_j).}
{}From the above equation its clear that a dipole whose ends points are
$X_1$ and $X_2$ respectively, the center of mass is sensitive to which
brane $i,j$ it ends. In fact the mass, $m_{ij}$,
of the string connecting the two
D6 branes, at ${\buildrel \rightarrow \over r}=
{\buildrel \rightarrow \over {r_i}}$ and
${\buildrel \rightarrow \over r}=
{\buildrel \rightarrow \over {r_j}}$,
can be easily determined from our M-theory picture. It is given
by the following integral over
(${\buildrel \rightarrow \over r}, x^7$) space:
\eqn\rxseven{m_{ij}=
 T_{M2}\int_{S_{ij}}
H_0^{1\over 2} H_1^{-{1\over 6}}H_2^{-{1\over 6}} H_3^{-{1\over
6}}\vert d{\buildrel \rightarrow \over r} \vert ~dx^7  \, ,} where
$T_{M2}$ is the tension of the M2 branes and $S_{ij}$ is the
sphere build by shrinking the (${\buildrel \rightarrow \over r},
x^7$)  torus at two points ${\buildrel \rightarrow \over {r_i}}$
and ${\buildrel \rightarrow \over {r_j}}$ as shown below:
\vskip.1in

\centerline{\epsfbox{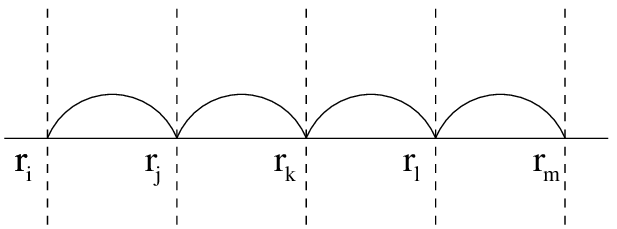}}\nobreak

\vskip.1in
\noindent We will assume that the
total mass of the string is distributed in a particular way when we
go to the LSQM approach. The center of mass $X_{(c)ij}$ is therefore
important.
Though its meaningful to write $X_{(c)ij}\equiv X_{ij}$,
we will
however use only $X_c$ to denote the center of mass. The full components
will be used only where we need to specify the exact details of interactions.

The change of basis is now readily calculable and its given by
\eqn\ChangeOfBasis {
    {\langle{e_i^a,e_j^a}\vert {X_c}\rangle} =
    e^{-\imath (e_i^a - e_j^a)
({1\over 2}(T^{-1}_i+T^{-1}_j) X_c)^A}
    \delta \left(\frac {e_i^a + e_j^a} 2 -
        \frac {((T^{-1}_i+T^{-1}_j) X_c)^a} 2\right),}
where the equality is upto a normalisation.

Now its again straightforward to use \aaProduct. The result can be
presented compactly as before \refs\KesYin:
\eqn\answer{
 (\Psi^i_j * \Phi^j_k)(X) =  \exp {\paren {\frac \imath 2
        \frac \pa {\pa{X'^{\mu}}}\Omega_{ij;jk}
         \frac \pa {\pa{X''^{\nu}} }}}
          \Psi^i_j(X')\Phi^j_k(X'')
          \Big\vert^{X'=S^{ik}_{ij}X,X''=S^{ik}_{jk}
          X} , }
where $\Omega_{ij;jk}$ can be given in terms of $T_i$'s as:
\eqn\defofomega{
\Omega_{ij;jk}= \left( {T^{-1}_i+T^{-1}_j \over 2} \right) ^{-1}J~
\left( {T^{-1}_j+
      T^{-1}_k \over 2}\right) ^{{\top}.{(-1)}}.}
The above equation is the precise mathematical formulation of non-Abelian
geometry. The noncommutativity parameter $\Omega_{ij;jk}$ is now
sensitive to which brane the dipole end points lie. The quantity $J$
 is the usual noncommutativity\foot{Multiple noncommutativity on the branes
were first discussed in \refs\RT, in connection with conifold geometry, and
in \refs\DolNap, for parallel branes in a slowly varying background field.}
(due to $B$ field) and $T_i$'s
are due to the non abelian nature. \defofomega\ therefore encodes this
intertwining clearly and there is no way to separate them. In terms of
string diagrams we need
\vskip.1in

\centerline{\epsfbox{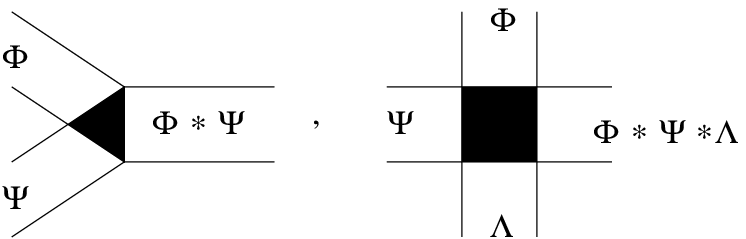}}\nobreak

\vskip.1in
\noindent where the basic triangle vertex and the next square vertex will be
used in the following
 section to understand some of the dynamics of the theory.
These vertices are meant to simplify the evaluations of the nonabelian
star product and have no other deeper significances.

\section{Some Applications}

So far we have developed the formalism of deformed product with
some relation to string theory. Even though the Taub-NUT setup has
its roots in string theory, its not very obvious how to embed the
nonabelian geometry in string theory. As discussed earlier,
stability is an issue here. Therefore there are two interesting
aspects of the problem: (a) non-trivial $H_{NSNS}$ background, and
(b) supersymmetry. Since $H_{NSNS}$ contributes to the
energy-momentum tensor we would need a curved background. For the
case when the $H_{NS}$ field was oriented in such a way that it is
constant along the D6 branes, the TN background in M-theory was
sufficient to explain the dynamics. As we saw in great detail
above, the M-theory  way of looking at it has an advantage. It
gives us an unified way to study many related theories which are
generated by choosing (in string theory) various orientations of
$B$ field along the brane \refs{\CDGR, \bergan, \berraj,\KesYin,
\seiwit }. The most generic background is of course the multi-TN
background with G-fluxes with or without M5 branes. Understanding
the dynamics of this background will give more detailed account of
all the theories mentioned and their product structure. Let us now
study some applications of the various scenarios discussed in the
earlier sections.

\subsection{Parallel Branes}

The information contained in $\Omega_{ij;jk}$ can be represented in terms
of dipole diagrams which can clarify the main idea of non-Abelian geometry.
Let us consider N D-branes labelled by $i,j,k,...$ and fix two points on
branes $i$ and $k$. Then to calculate the deformed product we need to
give weight to the triangle\foot{Observe that the two legs $ij$ and $jk$ of
the triangle
only provide spatial components from ${D\over 2}+1$ to $D$ or $1$ to ${D\over
2}$.
 Therefore to get
{\it all} the components we have to include the leg $ik$. This is also in
some sense the root of the ``cross'' product. From \ChangeOfBasis\ it is
clear that $X_c^a$ are correlated to $X_c^A$ via the matrices $T_{i,j,k}$ and
$J$ and vice-versa. There is of course no correlation between $aa$ and $AA$.
Therefore the noncommutativity parameter comes from full triangle $ijk$. And
since $j$ is summed over, all branes participate.}

\vskip.2in

\centerline{\epsfbox{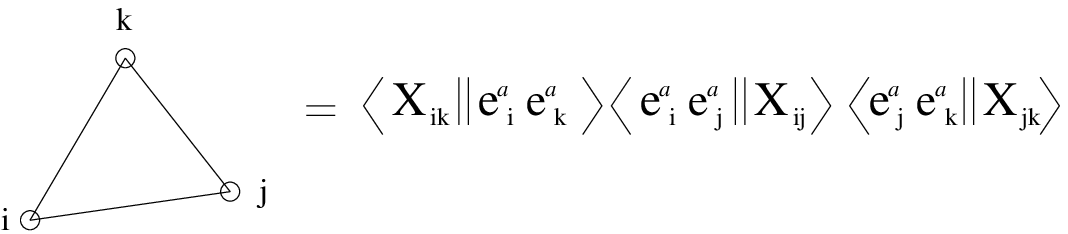}}\nobreak

\vskip.2in

\noindent where $j$ is any arbitrary point.
For the given points $i,k$ the orientation
of the triangle is:
$${\buildrel \longrightarrow \over {ij}}+
{\buildrel \longrightarrow \over {jk}}
+ {\buildrel \longrightarrow \over {ki}}=0.$$
Therefore the quadratic interactions are
given by the following diagram:

\vskip.2in

\centerline{\epsfbox{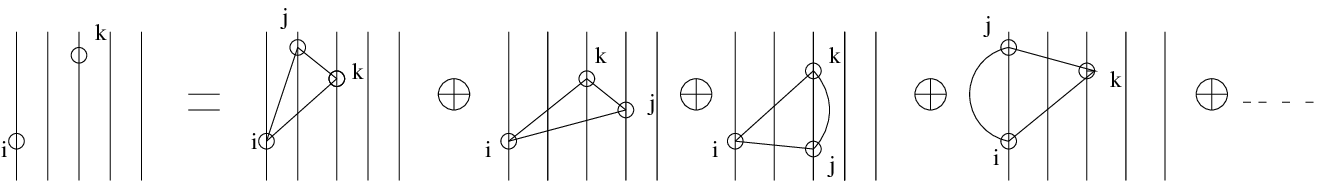}}\nobreak

\vskip.2in

$$~~~~~~~~~~~~~= \sum_j~e^{~[ \left( T^{-1}_i+T^{-1}_j \right) ^{-1}J~
\left( T^{-1}_j+
      T^{-1}_k \right) ^{{\top}.{(-1)}}]}$$
Let us now consider a case in which the wavefunctions are concentrated
near one of the brane labelled by $1$. In terms of the above formula we
now have\foot{
By $[~~]$ here we always mean that the terms are contracted by the
corresponding momenta of the wavefunctions appearing alongwith.
 For details see \refs\KesYin.
We also neglect the numerical factors.}
\eqn\nearone{
 \sum_{k=1}^N~ e^{~[(1+U_{1k})^{-1} \Omega_1 (1+U_{1k})^{{\top}.{(-1)}}]} =
e^{[\Omega_1]} \oplus ....}
where $U_{1k}= T_1T^{-1}_k$.

This would imply that near one of the brane (say $1$), the non-commutativity
parameter is given by  $\Omega_1$. This is consistent with naive expectation
\refs{\DolNap}. Here what we see that the non-abelian nature could in
principle contribute to $\Omega_1$. However as we shall show in the next
section, the matrix $U_{ij}$ depends upon the ratio of the two background
fields $\cF_{i,j}$.
For small variations of the fields $U_{ij}$'s  are
small. In terms of dipole diagrams we have

\vskip.2in

\centerline{\epsfbox{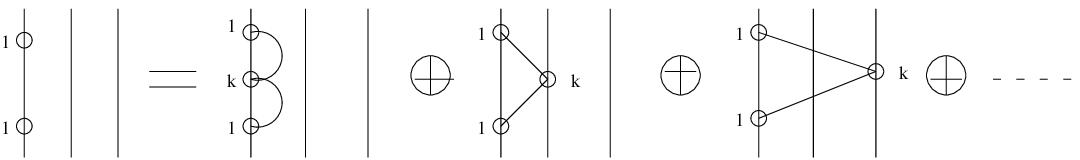}}\nobreak

\vskip.2in

\noindent Observe that the first diagram contributes as $\Omega_1$.

Let us see what are the fundamental dipole diagrams if we go to
higher order in interactions. Consider the following cubic interaction:
\eqn\cubicinter{
(\Phi * \Psi * \Lambda)^i_l =
(\Phi^i_j *_{ijk} \Psi^j_k) *_{ikl} \Lambda^k_l.}
Of course the above product is associative. Therefore it doesn't matter
which way we multiply.
It turns out that the explicit product involves the following terms:
$$\Omega_{ij;jk},~~\Omega_{jk;kl},~~\Omega_{ij;kl}$$
In terms of dipole diagrams this is:

\vskip.2in

\centerline{\epsfbox{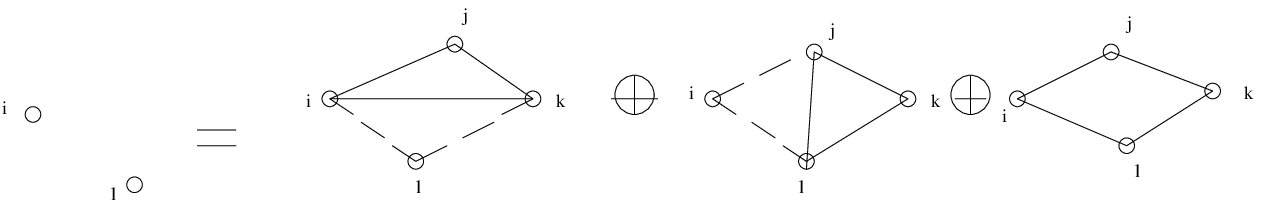}}\nobreak

\vskip.2in

\noindent As is obvious from above knowing the triangle and the quartic we can
predict all
higher order interactions.

Another related observation is to have two branes located at the two
opposite ends of a circle. For example let us have two parallel $D3$ branes
oriented along $X^{0,1,..3}$ and are on a circle $x^6$. Let us also put
gauge fields such that
\eqn\gaugfield{
\cF^1_{23}= \epsilon^{-\alpha}, ~~~~\cF^2_{23}=\epsilon^{-\beta}.}
This model can be related to the $D0-D4$ system studied in \refs\seiwit\ where
the $D4$ has a large amount of flux. For two $D3$ branes parallel to each
other and having some flux  the shift in the mode numbers
\eqn\shiftinmode{
\nu = {1\over \pi}(tan^{-1} \cF_1 - tan^{-1} \cF_2) = 0.}
(For equal and opposite fluxes the shift $\nu = 1$ \refs\KesYin). We now ask:
what is the deformation product for this theory?

The above model can be represented by an infinite array of $D3$ branes with
noncommutativity parameters $\Omega_1, \Omega_2, \Omega_1,..$.

\vskip.2in

\centerline{\epsfbox{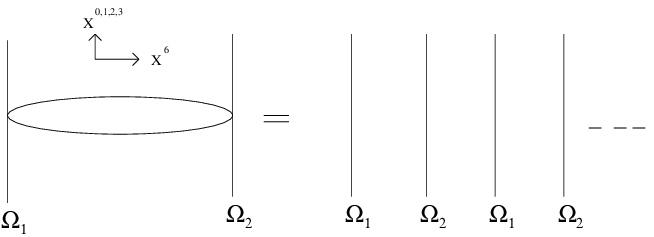}}\nobreak

\vskip.2in

 \noindent This would
mean that the infinite array of dipole diagrams can now be represented by
only two diagrams, implying, for example for the case $\alpha < \beta$,
\eqn\newcase{
\sum_k~ e^{~ [\Omega_{1k;k1}]}= e^{[\Omega_1]} \oplus e^{[4 \Omega_2]}}
where we have properly normalised the RHS.

\subsection{Brane-Antibrane theory}

So far we have been concentrating on scalar fields. Let us now take a
configuration of  branes and  antibranes stacked parallel and alternate to
each other. Let us denote them by $i,j,k,..$ where $i$ is a brane, $j$ is an
antibrane and so on. The $ii$ string will give vector multiplets (containing
vectors $A_{\mu}$ and scalars $\Phi$)
and so
would the $jj$ string. But the strings $ij, jk, kl$ etcs. will each give
rise to a complex tachyon:
${\cal T}_{ij}, {\cal T}_{ji} = {\cal T}^*_{ij}$.
Because of the
tachyons this system is unstable and it breaks all supersymmetry.

For the moment let us not bother about the instability of the system. We shall
return to this issue soon. Also for definiteness let us concentrate
on a system of $D5$ and $\bar{D5}$ labelled by $1$ and $2$ respectively.
The five branes are oriented along
$x^{0,1,..,5}$.

Along directions $x^{4,5}$ we switch on a gauge field $F_i$ and a $B$ field,
such that on each brane we have $\cF_i = F_i - B$. This will make the
theory on each brane non-commutative. For later use let us define the
following quantities:
\eqn\delupdown{
\cF_i- \cF_j \equiv \Delta,~~~~ {1\over 4}(\Omega_i - \Omega_j) \equiv
\nabla.}
Let us first consider the product:
$({\cal T}*\Phi)_{12}= {\cal T}_{12}* \Phi_{22}$. From our previous
discussions this will be defined by
\eqn\firstprod{ \Omega_{12;22} = (1+ U_{21})^{-1} \Omega_2 \, ,}
where we have already defined the matrix $U_{ij}$. It turns out that the
matrix $U_{ij}$ satisfy
\eqn\usatisfy{
\Omega_i = U_{ij} \Omega_j U^{\top}_{ij}}
as well as the cocycle condition
\eqn\cocycle{
U_{ij} U_{jk} = U_{ik}.}
The above two relations can be used, along with the fact that the
center of mass coordinate $X_c$'s commute among themselves, to rewrite
$U_{ij}$ in terms of the background fields $\cF_{i,j}$ as
\eqn\uijdefined{
U_{ij} = (1+2\Gamma \cF_i)^{-1} (1-2\Gamma \cF_j).}
This is the relation we want for extracting more information from
\firstprod. The matrix $\Gamma$ appearing in \uijdefined\ can be
related to $\Delta$ and $\nabla$ defined earlier via the following
relation:
\eqn\gamadefi{
\Gamma^{\top} - \Gamma + \Gamma \Delta \Gamma^{\top} = \nabla.}
The above relation exists as one can easily show that $\nabla \Delta$  can
never have eigenvalue $1$.

Before going any further we need to see how the tachyons behave in this
system. For this
let us study the system of $D5 - {\bar D5}$ with
fluxes on its world-volume in some details\foot{A somewhat similar
behavior of tachyons were studied in a different context for a system
of a D3 parallel to a D7 in the presence of non-primitive fluxes in \renata.}.
The reason why we took a system of $D5 - {\bar D5}$ is because this
configuration can be related to the case of a
$D5 - {\bar D5}$ brane wrapped on a vanishing two cycle of a conifold
\refs{\RT}. There is a non trivial $B$ flux on the two cycle, and since the
two cycle is of vanishing size the $B$ field is actually infinite. Similarly
there is also a gauge flux on the brane. It was conjectured in \refs{\DMUK}
that for some special choice of the background fields the system {\it is
supersymmetric} and the tachyon is massless.
This was eventually shown to be the case
in \refs{\RT}.

The quantization of open strings connecting a brane to an antibrane is
well-known.
In the NSR formalism, it is identical to the usual
quantization of open strings except that the GSO projection is
opposite to the usual one. Hence we keep the ``anti-GSO''
states. These include, at the lowest levels, a tachyon of $M^2 = -\half$.
In addition we find a set of massless fermions which are obtained by
dimensionally reducing a single 10-dimensional Majorana-Weyl fermion,
of opposite chirality to the usual GSO-projected one, down to $p+1$
dimensions.

In the case of interest here, this quantization is modified for two
reasons. First, there is a constant B-field, experienced by both the
brane and the antibrane, and also a world-volume field strength $F$ on
one of the pair. Second, the brane and antibrane are both wrapped
around a 2-cycle of vanishing size.

Let us work out the quantization of open strings joining a D5-brane to
a \Dbar 5-brane in the presence of fluxes. Let $b_1 = (F_1 - B)$ and
$b_2 = (F_2 - B)$, where $F_i$ are the worldvolume gauge fields on the
$i$th brane and $B$ is the constant spacetime $B$-field. Also, let $z
= x^4 + i x^5$. We have chosen to allow nonzero $F$ and $B$ values only
along the two directions $x^{4,5}$.

The boundary conditions are:
\eqn\boundcon{
\eqalign{
\left(\del_\sigma z + b_1\, \del_t z\right)_{\sigma = 0} &= 0\cr
\left(\del_\sigma z + b_2\, \del_t z\right)_{\sigma = \pi} &= 0}}
and a similar condition for $\bar z$ ( with $b_i \leftrightarrow
-b_i$).

Let us now write the mode expansion as:
\eqn\modeexp{ z = \sum_n A_{n + \nu}\, e^{(n + \nu)(t + i\sigma)}
+ \sum_n B_{n + \nu}\, e^{(n + \nu)(t - i\sigma)} \, .}
The first boundary condition yields
\eqn\firstbound{ A_{n + \nu} = B_{n + \nu}\,{1+ib_1\over 1-ib_1}
\, ,}
while the second one gives
\eqn\nueqn{ e^{2\pi i\nu} = {(1-ib_1)(1+ib_2)\over
(1+ib_1)(1-ib_2)}\, .}

Now recall that the 2-cycle on which $b_i$ are valued is of zero size,
which is the same as saying that the value of the field $b_i$ is
infinite, to keep constant flux. Thus we really need the above formula
for infinite $b_1,b_2$, except that they can each be separately
positive or negative infinity. Solving the above, one finds:
\eqn\nusolv{
\nu ={1\over 2\pi}\left( - {\rm tan}^{-1} {2b_1\over 1 - b_1^2}
 + {\rm tan}^{-1} {2b_2\over 1 - b_2^2} \right)\, .}
Now the relevant values are:
\eqn\relevvalues{
\eqalign{
 & {}^{\rm lim}_{b\rightarrow -\infty}~
{\rm tan}^{-1} {2b\over 1 - b^2} = 0 ,\cr & {}^{\rm
lim}_{b\rightarrow 0}~ {\rm tan}^{-1} {2b\over 1 - b^2} = \pi ,\cr
& {}^{\rm lim}_{b\rightarrow \infty}~ {\rm tan}^{-1} {2b\over 1 -
b^2} = 2\pi \, .}} These formulae are analogous to similar results
in Ref.\refs\seiwit\ for the D0-D4 system. It is interesting to
compare the two at this stage.  In the latter case, there are
altogether two pairs of directions over which a flux is allowed,
while we have only one. Also, in that problem there are DN strings
even in the absence of flux because the two branes have different
dimensions. While Ref.\refs\seiwit\ finds that a flux can make a
tachyon appear in a system that was BPS before, we will instead
find that a flux can make a tachyonic brane-antibrane system into
a BPS configuration.

Returning to the formula \nusolv, we can use \relevvalues\ to
evaluate it for the relevant possibilities
($\vert b_i\vert =0,\infty$ for each
$i$). The result is easily seen to be
\eqn\nuresult{
\nu = \half\Big( {\rm sign}(b_2) - {\rm sign}(b_1) \Big) }
where ${\rm sign}(b_i) = 0,\pm 1$.

Note that for our purposes, $b_1 = F_1 - B$, $b_2 = F_2 - B$. Hence
the case which we expect to be BPS comes about when $F_1=0$, so that
${\rm sign}(b_1) = -1$ while ${\rm sign}(b_2) = 1$, and $\nu=1$. On the
other hand, with $F_1=F_2=0$ we would find $\nu=0$ and this is the
case where we do expect a tachyon.

It only remains to find out the zero-point energy as a function of
$\nu$, in the NS sector (which is where the tachyon appeared, in the
absence of flux). We use:
\eqn\zeroptsum{ \sum_{n\geq 0}~ (n+\nu) = -{1\over 12}(6\nu^2
-6\nu +1)\, .}
The bosons along direction $x^{4,5}$ are quantised with mode numbers
$n+\nu$ and the fermions have mode numbers
$n \pm \vert \nu - {1\over 2}
\vert$. Thus the zero point energy of the system will be
\eqn\zeroptexp{ E = 2E(0) + E(\nu) + 2E(0) -E(0) -3E({1/2}) -
E(\vert \nu - 1/2\vert)\, .}
The first term comes from $x^{0,1,2,3}$, the second from $x^{4,5}$,
the third from $x^{6,7,8,9}$ and the fourth from the bosonic ghosts. The
remaining terms are fermionic contributions. Adding up all the
contributions, we get:
\eqn\eground{ E = -{1\over 2}\left(\vert \nu - {1\over 2} \vert +
{1\over 2}\right) \, .}
The case of no fluxes is $\nu = 0$ while the case of fluxes relevant to
fractional branes, as we argued above, is $\nu = 1$. From the above formula
we seem to find that both $\nu = 0$ and $\nu = 1$,
the
ground-state energy is $E=-\half$ and hence there is a tachyon.

However the actual result is more subtle because of the GSO
projection\foot{The discussion in the next two paragraphs were explained
to us by Sunil Mukhi. We thank him for a detailed discussion of the
tachyonic behaviour of this model. See also \bak\ for a 
somewhat related system.}.
At zero flux, along with the open string tachyon there is
always a massless state created by a world-sheet fermion (in the NS
sector) $\psi_{-\half}$.  This is in fact a spacetime scalar or vector
(depending on whether the fermion mode has an index transverse to the
brane or along the brane). Now when there is flux, this mode (for the
directions along which the flux is present) becomes
$\psi_{-\vert \nu-\half\vert}$. Thus the corresponding state has energy
\eqn\mukhi{E=
-\half\left(\vert \nu-\half\vert + \half
\right) + \vert \nu-\half\vert =
\half\left(\vert \nu-\half \vert  -\half\right).}
Thus altogether we have a pair of low-lying states, one of energy
$-\half(\vert \nu-\half\vert  + \half)$ and the other of energy
$\half(\vert \nu-\half\vert  -\half)$. At
$\nu=0$ these states have energies
$-\half,0$ respectively, and at $\nu=1$ they also have energies
$-\half,0$. But if we tune $\nu$ continuously from 0 to 1, we find
that at $\nu=\half$ the states become degenerate in energy, with both
having $E=-{1\over 4}$. It turns out that at this point the two states
cross each other.

To see this more explicitly, observe that the energies of the pair of
states can equivalently be written $-\half\nu$ and $\half(1-\nu)$ for
all $\nu$, without any mod sign. In this way of writing it, the
energies vary smoothly with $\nu$. These expressions, and not the
earlier ones involving modulus signs, are the correct ones if we want
to follow the evolution of a given state (with a given sign under GSO
projection) as $\nu$ varies. Now we see that the tachyon at $\nu=0$
becomes massless at $\nu=1$. On the other hand the massless state at
$\nu=0$ becomes tachyonic at $\nu=1$. But since we are in a sector
with anti-GSO projection, the latter state is projected out for any
$\nu$! The physical (anti-GSO) state, which is tachyonic at $\nu=0$
really does become massless at $\nu=1$. Thus we have shown that the
tachyon disappears in the presence of flux, as desired. In the figure
below we sketch the behavior of the tachyon for this system:

\vskip.1in

\centerline{\epsfbox{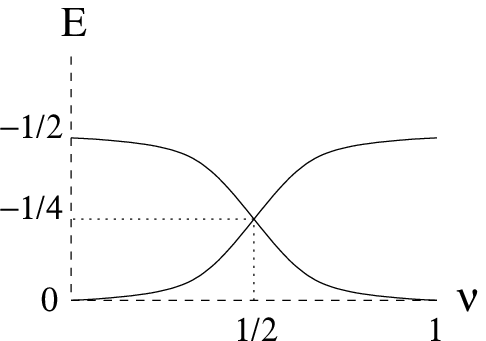}}\nobreak

\vskip.1in
\noindent There are two subtleties in the problem
which we shall be ignoring. First,
the $B$ and the $F$ fields on the two cycle are not constant so
we cannot actually use the whole arguments of non-abelian geometry. But we
shall take a simplified case in which the fields are actually constant. We
will return to the issue on nonconstant fields on the brane in the next
section.
The second subtlety has to do with the spherical cycle. We shall take the
brane and the antibrane to be wrapped on a toroidal cycle.
For vanishing cycle
we will take the limit in which the fields approach infinity such that their
ratio is finite.
Therefore, for finite size of the cycle,  $U_{21}$ is finite and small and
$\Omega_{12;22} = \Omega_2 + {\cal O}(U)$.
Thus the multiplication rule is
\eqn\tachmult{
{\cal T}_{12}  * {\Phi}_{22} = {\cal T}_{12} *_2  {\Phi}_{22} + {\cal O}(U).}
Since both $\cF_1/ \cF_2$ and $\cF_2/ \cF_1$ can be made finite, $U_{12}$
is also finite. This in turn implies:
\eqn\implies{
{\cal T}_{21}  * {\Phi}_{11} = {\cal T}_{21} *_1 {\Phi}_{11} + {\cal O}(U).}
Question now is what should be ${\cal T}_{12}* {\cal T}_{21}$? Here it would
seem that either of $*_1$ or $*_2$ could suffice. To fix this let us go to
higher order interactions: $({\cal T}* \Phi *{\cal T})_{11} =
{\cal T}_{12}*{\Phi}_{22}*{\cal T}_{21}$. From the dipole diagrams in the
previous section we have seen that this is uniformally generated by:
\eqn\dipnowhelps{\eqalign {
[\Omega_{12;22}]\oplus  [\Omega_{22;21}] \oplus
[\Omega_{12;21}] =& [(1+ U_{21})^{-1} \Omega_2]
~~ \oplus ~~[\Omega_2 (1+ U_{21})^{{\top}.{(-1)}}] \cr
 & \oplus  [(1+ U_{21})^{-1} \Omega_2 ~(1+ U_{21})^{{\top}.{(-1)}}]}}
implying that the product rule here could possibly be
${\cal T}_{12}*_2{\cal T}_{21}$.
And similar for other cases. The dipole diagrams are:

\vskip.2in

\centerline{\epsfbox{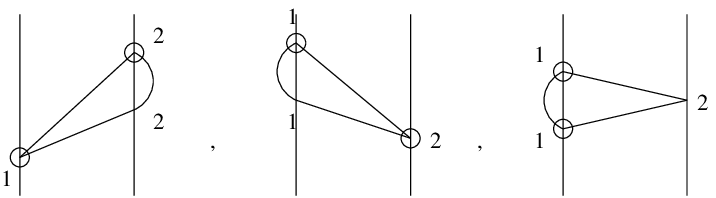}}\nobreak

\vskip.2in

There could be other ways of writing the
deformation product between the fields but the above procedure tells us that
a consistent approximation is\foot{Since the triangle has orientation, the
apparent ambiguity arising due to complex conjugation isn't there.}
\eqn\consappro{
\Psi_{ij} *_{ijk} \Lambda_{jk} ~ = ~ \Psi_{ij} *_j \Lambda_{jk} +
{\cal O}(U)}
for the specific case of this problem. $\Psi$ and $\Lambda$ could be
${\cal T}$, $\Phi$ or gauge fields.
Thus we see that \answer\ reproduces all the
known results and also shows what are the possible corrections. These
corrections are basically because of the intertwining of the usual *-product
and the non-abelian space.

These results can be easily extended for the case of nonbps branes. Consider
two $D6$ branes in type IIB theory. These are unstable branes and there are
now four tachyons in the system. In the presence of $\cF_i$ fields on the
branes one can use \answer\ to determine the product rules for multiplying
tachyons in this theory.

\subsection{Branes with $H_{NSNS}$}

So far we have considered the case when the $G$ fluxes were decomposed over
the ASD harmonic two form $\omega$ in such a way that the components along
the D6 brane(s) are constant. We can now consider the case when the $G$
flux is a very generic function of space coordinates. From D6 brane point of
view there are two interesting cases now:

(a) There is a non-trivial $H_{NS}$ field on the world-volume of
D6.

(b) There is a non-trivial $H_{NS}$ vanishing at the brane location.

\noindent Both these cases can actually
arise in the {\it hybrid} theory setup that we
described earlier. However we have to be a little careful now. The two D6
branes are oriented along $x^{0, 1, 2, ..., 6}$ and along
$x^{0, 1, ... , 4, 8, 9}$. In the hybrid theory that we describe we have,
among other three-form fields in type IIA theory, the $B$ field $B_{16}$
which have both its legs on one D6 and only one leg along the other D6.
This gives us a hint to use the same framework for the case when
we have a $B$ field which is (say) $B_{89}$. In terms of M-theory we need
a threeform flux $C_{789}$ in the conifold background. The above choice of
$B$ field immediately tells us that we have a pinned brane like scenario
on the other D6 (because the $B_{89}$ field is completely orthogonal to the
D6). In the pinned brane scenario we saw in the earlier sections that the
hypermultiplets pick up masses, and the mass is determined by the
asymptotic value of fluxes. What would this imply for the other brane? On the
other brane we see a $B$ field with both legs on the brane. But some of the
scalars that have become massive due to the first brane continue to be massive
on the other brane also. Furthermore since the $B$ field would vary
non-trivially on the orthogonal directions of the first brane as a result
we will see a {\it non-constant} $B$ field on the other D6 brane, i.e a
$H_{NS}$ field! Thus the conifold like geometry that we discussed in earlier
sections can in fact also allow us to study this scenario.

Question now is whether we could say something about any kind of
star product here? For the hybrid brane scenario, when we have
non-zero $H_{NS}$ on the other D6 brane, the open string
fluctuations are goverened by an additional term on the
world-sheet $\Sigma$ (discussed also in a slightly different
context in \corshi): \eqn\addte{ {i\over 3 \pi \alpha'}
\int_{\Sigma}~H_{abc}~X^a ~dX^b \wedge dX^c \, ,} where $H_{abc}$
is the threefrom on the D6 brane. There are of course the usual
terms on the world-sheet that we ignore here. This theory also
have a star product governed by (in some limits) the inverse of
${\tilde {\cal F}}_{ab} \equiv {\cal F}_{ab} + {1\over 3} H_{abc}
x^c$, where ${\cal F}_{ab}$ is used in the previous sections to
evaluate the Moyal product. If we ignore, for the time being the
other D6 brane giving rise to the pinned brane scenario, we would
have a similar TN background with fluxes. However the G-fluxes now
needed is {\it not} the one discussed in \gfluxch\ but rather a
more involved one as \eqn\gflnowis{{G \over 2\pi} = \sum_{i=1}^N
{\cal F}_i \wedge \omega_i + dz \wedge \gamma_1 + d{\bar z} \wedge
\gamma_2 \, ,} where $z, {\bar z}$ are one-forms in this
background (with some specific choice of complex
structure)\foot{These one forms parametrise a $T^2$ in the
conifold geometry. This is the same $T^2$ on which by making two
T-dualities we could get a brane box configuration in type IIB
theory.} and $\gamma_{1,2}$ as the three-forms that eventually
determine ${\tilde{\cal F}}$. This choice of G-flux is now used to
determine the star product for the system following the analysis
presented in \cita. It turns out that this star product is in fact
non-associative and the non-associativity is given by (first shown
in \refs\corshi): \eqn\noass{ (\Phi \star \Psi) \star \Lambda -
\Phi \star (\Psi \star \Lambda) = (\gamma_1+\gamma_2)_{abc}
\lambda^{abc}} where $\lambda^{abc}$ is in general non-zero and is
a function of the fields $\Phi, \Psi, \Lambda$ as discussed in
\corshi. The non-associative interaction is evident directly from
the BI action of the D6 brane written in terms of $\sqrt{{\rm
det}~(g + \gamma \cdot x + ..)}$ where $\gamma = \gamma_1 +
\gamma_2$ and $x$ is the coordinate on D6. Notice also that we
haven't put any conditions on $\gamma_{1,2}$. Supersymmetry would
require the minimal constraint: $\gamma_2 = \ast ~\gamma_1$.
However the fact that the algebra is non-associative hinders any
nice description of the system.

This situation could improve if we consider  the case when the $H$
field vanishes at the point where one of the D6 is placed\foot{A
similar configuration is recently been shown to occur (with D3
replacing the D6) in a parallellizable plane wave background
\refs\jabsad. It will be interesting to find a connection between
the two scenarios.}. Therefore a supersymmetic configuration with
the choice of G-flux \gflnowis\ is now further constrained. The
generic behavior, in the absence of field strength, is that
$\gamma_1$ should at least be anti self-dual. This ASD property
changes to \eqn\consg{\ast \gamma_1 = - \gamma_1 + \sum_{n}~a_n
r^n \, ,}
where $r$ is the coordinate orthogonal to
 the D6 brane. Because of the other D6 brane which is pinned to the
origin $r = 0$, our D6 brane will also remain fixed at the origin and therefore
we might naively expect the ASD property to be violated as
$\ast \gamma_1 = - \gamma_1 + a_0$. Making $a_0 = 0$ will give an exact ASD
condition on $\gamma_1$ such that the background three-form on the brane
$\gamma_1 + \gamma_2 = 0$. This would also mean that \noass\ is no longer
the case and the algebra will tend to be associative even though we have a
non zero three-form parallel to the brane. Observe that the pinning of the
branes in this case is very important. If we allow the un-pinning effect in
the conifold setup, then probably associativity for this case will be
broken. This needs to be verified.

There is yet another extension of the above discussion that could in principle
arise in the hybrid theory. Till now we saw two interesting cases: (a) Dipole
plus noncommutative theory, and (b) Pinned brane theory plus noncommutative
theory. We could extend this scenario further by taking the case of
rank six $B$ field on one of the D6 brane so that we have (for example)
the following components: $B_{12}, B_{34}, B_{89}$. For the other D6
oriented along $x^{0, 1, ..., 6}$ there is a noncommutative algebra along
directions $x^{0, 1,2 ,3, 4}$ but also some possibility of pinning
due to $B_{89}$ (whether or not there is a real pinning needs to be
verified by sugra analysis). On the other hand the second D6 sees a complete
noncommutative algebra now. We are assuming for simplicity that there is no
$H$ field and also all the $B$ fields have equal magnitudes. Viewing now the
conifold as an ALE fibration over a {\it constant} slice of the $P^1$ base
the value of $g(r)$ can be estimated for this case using the generic rule
given in \funcgras. The result is\foot{Done in collaboration with Govindan
Rajesh.}
\eqn\ranksixg{g(r) = {r~(1+{\rm sec}~\alpha)^2~
{\rm exp}\left(2\sqrt{2r + R~{\rm sec}^3
\alpha \over 2r ~{\rm sec}^2 \alpha + R~{\rm sec}^3 \alpha} -
{2\over {\rm sec}~
\alpha} \right) \over
(1+ {\rm sec}^2 \alpha) r + {\rm sec}^3 \alpha ~R +
\sqrt{(2r ~{\rm sec}^2 \alpha + R~{\rm sec}^3 \alpha) (2r + R~{\rm sec}^3
\alpha)}}.}
In deriving this we have ignored the effect of stuck D0 branes on the D6 brane
as mentioned in \refs\cita. Taking the other slice of the conifold we have
a rank four $B$ field on the D6 with an additional $B$ field orthogonal to
it. The detailed dynamics of this system is complicated which we hope to
tackle in near future.

\subsection{Theory near multi-Conifold type geometry}

So far we have discussed examples with single centered Taub-NUT geometry and
multi centered TN geometry. We also saw that a $P^1$ fibration of the
TN or ALE space gives rise to a conifold type geometry. Question now is to
understand what happens if we have the multi TN fibered over a $P^1$ base.
The family of resolved ALE spaces with $A_{n-1}$ singularities is given by
the equation
\eqn\resale{ z_1^2 + z_2^2 + \prod_{i=1}^n ~(z_3 - \mu_i) = 0}
where $z_i$ are the coordinates of a $C^3$ and the $S^2$ mentioned in the
earlier sections are related to $\mu_i$ as $S_{ij} \leftrightarrow
\mu_i - \mu_j$. The full $A_{n-1}$ symmetry is realised when all the $S^2$
shrink to zero size. From here it is now clear that the multi conifold
geometry can be realised when we consider the following fibration
\eqn\conifib{z_1^2 + z_2^2 + \prod_{i=1}^n~(z_3^2 + z_4^2 - \mu_i) = 0.}
Over the curve $z_1 = z_2 = 0$ we have the ALE $A_{n-1}$ transversal fiber.
The above consideration opens up yet another set of theories that could be
studied in M-theory on a multi conifold type of background in the presence
of fluxes and probed by a M5 brane background.

These singularities which have been discussed earlier in \refs\sadovafa,
can actually be related to the quotient of a conifold for the case when
$\mu_i = 0$. This quotienting is a simple ${\bf Z}_n$ action \refs\angle\
defined on the equation for the conifold written as
\eqn\coniwri{ XY = Z^n W^n}
where $X,Y,Z,W$ can be easily related to $z_i$ of \conifib. We could even
extend this further by taking the quotient operation as ${\bf Z}_m \times
{\bf Z}_n$. This geometry has also been discussed in \sadovafa\ and takes the
form
\eqn\takesthe{ X^m Y^m = Z^n W^n}
where we have again used the same coordinates. All these singularities are
of course related to ALE fibration over some base. However we could
go beyond that by taking a singularity for which powers of $Z,W$ are
different\foot{For example manifolds of the form $XY = Z^m W^n$. As discussed
in \refs\angle\ these manifolds can actually be generated in type IIB theory
{}from the T-dual picture where we have intersecting set of NS5 branes. Putting
$m$ number of NS5 branes along one direction and $n$ number along the
orthogonal direction can in fact generate this kind of singularity
via one T-duality.}.
This is clearly not related to  any quotient of a
conifold and might fall slightly off the main line of thought that we are
pursuing here. Nevertheless it will be interesting to see any such
connection.

All these M-theory configurations can be studied in type IIA also.
In the type IIA side, as before, there are two possibilities. We could
either come down via $x^1$ or via $x^7$ - the conifold cycle. The first case
is simply the reduced conifold geometry whereas the second case is a set
of $n$ orthogonal D6 branes with fluxes on its world volume. The hybrid
theory discussion that we had earlier will now be more involved because
we could now have a situation where the noncommutative geometry on one
set of parallel D6 branes could be converted to a nonabelian geometry!
Clearly now the star product is more involved because of this. But thats
not the end of the story. We saw in the conifold setup that we could have
dipole theory on the other brane. When we allow different fluxes on different
D6 then there is a possibility to have {\it different} dipole lengths on the
world volumes. The dipole star product will now have to mix somehow with the
nonabelian nature of the D6 branes. This theory has never been studied
before. It will be interesting to see whether a stable configuration can be
realised in this setup.

Before we end this section we should point out that this geometrical
construction can be extended even further by including the $D_n$ and
$E_n$ kind of singularities. First let us look back again at the TN scenario.
We discussed the following structures:
$${\rm TN}~~\to ~~{\rm Multi~~TN.}$$ We will
also consider this from ten dimensional
perspective. The above singularity is just the $A_n$ series of Kleinian
singularities discussed in the literature. The TN that we discussed earlier
was oriented along $x^{7,8,9, 10}$ (with $x^1$ as the M-theory circle).
Combining these to form $z^1, z^2$ in the obvious way, the $A_n$ singularities
are generated by action $\Gamma$ on $C^2/\Gamma$ with the action
being via a matrix with diagonal entries $e^{\pm{2i\pi\over n+1}}$. This
matrix, and together with an off-diagonal matrix with entries $i$ will
now generate the $D_n$ singularities. From M-theory point of view the
wrapped M2 branes should now reproduce the  intersection matrix of
$D_n$ kind of singularities. Such a scenario is possible
when we have a Atiyah-Hitchin kind of singularity on top of a TN
singularity, at least locally.
Analysis of this background in the presence of fluxes
is subtle and a fully compact
example (with all local charges cancelled) leads us to manifolds that are
generically non-K\"ahler \refs\bbdg.  Section 2.6 of \refs\bbdg\ discusses the
full implications of a $D_4$ singularity and therefore we refer the reader
to that. The $E_n$ singularities now fall under the binary
{\rm tetrahedral}, {\rm Octahedral}
and {\rm Icosahedral}
groups. The adjoining matrices in the three cases
are given by
\eqn\icaosa{\eqalign{&
T = {1\over \sqrt{2}}{\pmatrix{\epsilon^7 & \epsilon^7 \cr
\epsilon^5 & \epsilon}},~~~~~ O = {\pmatrix{\epsilon &0 \cr 0 & \epsilon}}\cr
& I_1 = {\pmatrix{-\eta^3 & 0 \cr 0 & - \eta^2}},~~~~ I_2 =
{1 \over \eta^2 - \eta^3}{\pmatrix{\eta + \eta^4 & 1 \cr 1 &
-(\eta + \eta^4)}}}}
where $\epsilon$ is the primitive eighth root of unity and $\eta$ is a fifth
root of unity. The tetrahedral group is generated by $T$ and the
octahedral group is generated by taking the elements of $T$ and adjoining them
with the elements of $O$. The icosahedral group is generated by $I_1$ and
$I_2$.

However this is not enough. There is still more to the above classification.
It turns out that the above set of singularities are related to the
{\it chiral} ground rings\foot{These are the polynomial rings that form a subclass of the infinitely
many extra states of a $c = 1$ string theory with non-standard ghost number. See \ceqlone\ for details.
The $A_n$ transition that we discussed above, for example, can be thought of as a polynomial ring in three variables that 
defines a complex variety in ${\bf C}^3$ at the self dual point of radius ${n \over  \sqrt{2}}$ in $c = 1$ theory.  
The non-chiral rings on the other hand are the {\it quantum} ground rings, in the language of \ceqlone. They are 
constructed by combining the left and the right movers in the $c = 1$ closed string sector. Since this aspect is 
somewhat orthogonal to the main line of thought followed in the paper, we refer the readers to Witten's work on 
the ground ring in $c = 1$ theory \ceqlone\ for further details.}
of a $c = 1$ string theory at the self-dual radius
\refs{\ceqlone, \ghomuk}. An immediate question would be:
what about the non-chiral rings? As discussed
in \refs\ghomuk, the non-chiral rings are responsible for the other kind of
structures that we studied here, namely $${\rm Conifold} ~~ \to ~~
{\rm Multi~Conifold.}$$ The above structure is nothing but the $A_n$
series for the non-chiral ring. Therefore one would ask about the
other $A-D-E$ singularities\foot{ The $E_n$ singularities are disconnected from the usual $c = 1$ CFT sector. And therefore the 
analysis of these are rather involved.}. 
Clearly many things can be said now, but
we will not pursue this here anymore and leave the rest of the discussions
for future work. The connection of our work to $c = 1$ string theory
provides yet another unified picture to view the whole dynamics.

\section{Discussions}

In this paper we followed two unified themes. First,
there is a unification at the
level of underlying gravitational solution. We start with a Taub-NUT space
with background three-from fluxes in M-theory. When this background is
fibered over a $P^1$ in some specific way we get a conifold like geometry
with fluxes. This gives us a way to interpolate between two different
backgrounds. Furthermore the Taub-NUT space could be made multi-centered.
Fibering this over a $P^1$ now gives us a multi-conifold like geometry.
Therefore in terms of geometry we have the following interpolating scenarios:

\vskip.15in

\centerline{\epsfbox{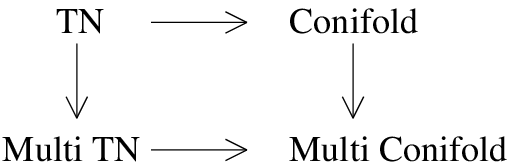}}\nobreak

\vskip.15in

\noindent Secondly, there is a unification at the level of theories probing
these backgrounds. All these theories appear in M-theory
{}from a configuration of branes, fluxes and geometry. In fact the
geometry (with fluxes) by themselves is responsible to generate
either noncommutative, nonabelian, or even hybrid theory depending on
specific configuration. When M5 branes probe this background, pinned
brane theory or dipole theory can be studied. This aspect is interesting
because many different theories are now different guise of one
unified framework in M-theory. We also pointed out that the above scenario
could be extended even further
if we allow non-trivial $H_{NS}$ fields on the
branes. A simple analysis showed that the star product becomes non-associative
and therefore this would not be physically relevant. However when the
$H_{NS}$ field vanishes on the brane (but no where else) then many interesting
things can be said.

All the above theories studied are static in nature. We could ask whether
it is also possible to study theories that have some inherent time
dependences. Some aspect of this have been dealt recently in
\refs{\savhashi,\cita}\foot{In \refs\savhashi\ explicit holographic dual
for a time dependent closed string background was presented. The
noncommutativity parameter was found to be non constant and the theory
on the brane was shown to be decoupled from the bulk. For our case here,
since we have taken D6 branes, the theory however is never decoupled.}.
Defining $x^\pm$ in the usual way, the metric in
M-theory is defined in terms of two harmonic functions $H_{1,2}$,
 such that they are related by
\eqn\honehtwo{ H_1 - H_2 = \left({x^+\over R_o}\right)^2\, ,}
where $R_o$ is the radius of the TN (the relation between $R_o$
and $R$ mentioned in this paper is given by $R =
g_s~l_s^2/R_o$). The explicit derivation of the above formula
\honehtwo\ is given in \refs\cita\ and so we refer the reader to
that for further details. To progress further along the generic
idea presented in this paper, we need the value of $g(r)$. This
can be derived using the generic rule given in \funcgras\ and the
steps mentioned in \refs\cita\ assuming the two form $\omega$ to
be normalisable and harmonic for a {\it fixed} $x^+$. In terms of
the variables defined above this is given by: \eqn\grtime{g(r) =
\left(1+\sqrt{1+\left({x^+\over R_o}\right)^2} \right)^2~ {2r
\over \left( \sqrt{R + 2r} + \sqrt{R + 2r~ \left( 1 + \left(
{x^+\over R_o}\right)^2\right)}\right)^2}\, ,} where we have again
measured $g(r)$ wrt $2\pi l_p^2$. Some aspects of this time
dependent background has been addressed in \refs\cita\ using the
arguments of duality chasing. Since this example also falls in the
description that we have given, broadens the horizon of the
application of our general framework to time dependent cases also.
It will now be interesting to see whether we could say something
more about the conifold, multi TN and its generalisations from
here. Clearly we have just scratched the surface and many more
details need to be filled in to get a deeper understanding of the
dynamics of oriented $B$-fields. We hope to address this in near
future.

\vskip.2in

\noindent {\bf Acknowledgements}

\noindent We would like to thank Sunil Mukhi for an explanation of the
tachyonic behaviour of a brane anti-brane system and Govindan Rajesh
for helping us calculate some of the properties of a Taub-NUT geometry.
We would also like
to acknowledge useful conversations with
Ori Ganor, Daniel Robbins,
M. M. Sheikh-Jabbari, Savdeep Sethi, and Zheng Yin
on  various aspects of these theories.
The work of K.D is supported in part by a David and Lucile Packard Foundation
Fellowship 2000-13856. The work of M.S is supported in part by DOE grant
number DE-AC03-76SF00515.

\listrefs
\end

\ \medskip
\begintable
{\it Theories}|{\it SUSY}|{\it Product rule}:~$\Phi*\Psi(X)$\eltt
Pinned branes|$\cN=2$|$\Phi(X) \Psi(X)$\elt
Dipole theory|$\cN=2$|$e^{{1\over 2}(L_1^i{\del\over \del X''^i}
-L_2^j{\del\over \del X'^j})}~\Phi(X')\Psi(X'')\vert^{X'=X''=X}$\elt
Noncomm. geometry|$\cN=4$|$e^{\paren {\frac \imath 2
         \frac \pa {\pa{X'^{\mu}}}\Omega^{\mu\nu}
         \frac \pa {\pa{X''^{\nu}} }}}~
    \Phi(X') \Psi(X'') \vert^{X'' = X' = X}$\elt
Non-Abelian geometry|$\cN=0$|
$e^{\paren {\frac \imath 2
        \frac \pa {\pa{X'^{\mu}}}\Omega_{ij;jk}^{\mu\nu}
         \frac \pa {\pa{X''^{\nu}} }}}~
          \Phi^i_j(X')\Psi^j_k(X'')\vert^{X'=S^{ik}_{ij}X}_{X''=S^{ik}_{jk}
          X}$
\endtable